# Synthetic spin orbit interaction for Majorana devices

M.M. Desjardins[1,†], L.C. Contamin[1,†], M.R. Delbecq[1], M.C Dartiailh[1], L.E. Bruhat[1], T. Cubaynes[1], J.J. Viennot[2], F. Mallet[1], S. Rohart[3], A. Thiaville[3], A. Cottet[1] and T. Kontos[1]*

[1] *Laboratoire Pierre Aigrain, Ecole Normale Supérieure-PSL Research University, CNRS, Université Pierre et Marie Curie-Sorbonne Universités, Université Paris Diderot-Sorbonne Paris Cité, 24 rue Lhomond, 75231 Paris Cedex 05, France*

[2] *Institut Néel, CNRS, 25 rue des Martyrs, 38042 Grenoble cedex 9, France*

[3]*Laboratoire de Physique des Solides, Université Paris-Sud et CNRS, Bât. 510, 91405 Orsay Cedex*

*To whom correspondence should be addressed: kontos@lpa.ens.fr

† These authors contributed equally to this work

**The interplay of superconductivity with a non-trivial spin texture holds promises for the engineering of non-abelian Majorana quasi-particles. A wide class of systems expected to exhibit exotic correlations are based on nanoscale conductors with strong spin-orbit interaction, subject to a strong external magnetic field. The strength of the spin-orbit coupling is a crucial parameter for the topological protection of Majorana modes as it forbids other trivial excitations at low energy[1,2]. The spin-orbit interaction is in principle intrinsic to a material. As a consequence, experimental efforts have been recently focused on semiconducting nano-conductors or spin-active atomic chains contacted to a superconductor[3,4,5,6,7]. Alternatively, we show how both a spin-orbit and a Zeeman effect can be autonomously induced by using a magnetic texture coupled to any low dimensional conductor, here a carbon nanotube. Transport spectroscopy through superconducting contacts reveals oscillations of Andreev like states under a change of the magnetic texture. These oscillations are well accounted for by a scattering**

**theory and are absent in a control device with no magnetic texture. A large synthetic spin-orbit energy of about 1.1 meV, larger than the intrinsic spin orbit energy in many other platforms, is directly derived from the number of oscillations. Furthermore, a robust zero energy state, the hallmark of devices hosting localized Majorana modes, emerges at zero magnetic field. Our findings synthetize all the features for the emergence of Majorana modes at zero magnetic field in a controlled, local and autonomous fashion. It could be used for advanced experiments, including microwave spectroscopy and braiding operations, which are at the heart of new schemes of topological quantum computation.**

Hybrid superconducting nanoscale systems are appealing for fundamental studies on how superconducting correlations are altered when the spin and the motion of electrons are coupled[3,4,5,6,7]. When a superconductor is connected to a nanoscale conductor which hosts few coherent channels, the superconducting correlations give rise to a discrete spectrum inside the superconducting gap, called Andreev bound states. Whereas these states are now understood exquisitely in conventional setups, their fate in topologically non-trivial materials is presently the subject of an intense activity. One particularly striking feature of those systems is the possible emergence of Majorana modes.

Engineering robust Majorana modes in hybrid quantum devices requires non-collinear spin orientations and superconductivity. This can be achieved through the interplay of a homogeneous magnetic field and spin-orbit coupling. Alternatively, it has been theoretically proposed to synthesize these ingredients at the microscopic level by mimicking both a spin orbit coupling and a Zeeman splitting using a magnetic texture[8,9,10,11,12,13,14,15], as sketched in Fig. 1a. Here, we demonstrate such a device using a single wall carbon nanotube (our 1D conductor) connected to two superconducting electrodes and coupled to a proximal magnetically textured gate. Within the superconducting gap, we observe Andreev-like States (ALSs) whose energy oscillates as a function of the external magnetic field. This is direct evidence that we have induced a large synthetic spin orbit interaction in the nanotube and observed its interplay with spinful Andreev bound states. We finally observe a zero bias conductance peak stable in magnetic field, which is compatible with the emergence of Majorana modes in our setup.

A magnetic field oscillating in space is mathematically equivalent to a Rashba spin orbit interaction combined with an effective constant magnetic field[8,9,10,11,12,13,14,15]. The period of the oscillations, $\lambda$, sets the spin orbit energy and their amplitude sets the magnitude of the effective field. Several methods have been suggested to induce a suitable oscillating field, such as the spin helix arising from nuclear spins interacting through a RKKY interaction[10] or specific patterns of micromagnets[9,11,12,13]. Importantly, such a scheme is expected to be robust to imperfections of the magnetic texture[13,16]. In this work, we elaborate on these theoretical approaches and use a single magnetic material whose domain structure gives rise to a cycloidal-like dipolar field along the nanotube, as depicted in Fig. 1a. Our device is shown in Fig. 1b and 1e. A single wall nanotube is stamped onto a magnetic $[Co(2.0nm)/Pt(1.2nm)]_{10}/AlOx$ multilayer bottom gate (see Extended data). The Co/Pt is expected to have a small pitch and an out of plane anisotropy, giving rise to several domains over the length of the nanotube, with a strong stray field of about 0.4 T. The Magnetic Force Microscope (MFM) picture shown in Fig. 1e evidences magnetic domains in the bottom gate, which have a typical size of about 100 nm (see Supplementary Information section 5). An external magnetic field $B_{ext}$ changes the magnetic structure and can therefore reveal the existence of the synthetic spin orbit interaction.

Superconducting correlations are induced by connecting the nanotube to two Nb(40nm)/Pd(4nm) superconducting electrodes. We address the discrete spectrum induced by the superconductor by transport spectroscopy[3,4,5,6,7]. The typical measurement of the differential conductance G as a function of source-drain bias $V_{sd}$ is shown in Fig. 1c. The conductance displays a well-defined energy gap of about 550 μeV containing two peaks, symmetric with respect to zero bias. These two peaks signal ALSs arising from superconducting correlations. As sketched in Fig. 1d, our measurements are equivalent at low energy to conventional tunnel experiments as a consequence of the finite density of states at the Fermi energy in one of the two superconducting contacts (contact 2). The global shape of the conductance curve is well accounted for by the quasi-classical description of superconductivity in the electrodes, based on Usadel equations and reveals that contact 1 displays a well-defined superconducting hard gap. The large subgap slope is shown to arise mainly from a residual pair-breaking in the superconductor (see Extended data). The ratio between the high bias conductance and the zero bias conductance which measures the "hardness" of the gap is of about 45 which compares favourably with the recently reported figures in semiconducting nanowires[5].

One of the main findings of our experiments is displayed in Fig. 2b. In this colorscale map of G as a function of $V_{sd}$ and the external magnetic field $B_{ext}$, we observe the evolution of the ALSs under an external magnetic field. Strikingly, they display oscillations with a period of about 600 mT (+/- 10% from one magnetic sweep to another).

We can resolve up to three oscillations around the mean energy of 220 μeV, together with the expected slow reduction of the superconducting gap as shown in Fig. 3a. Such a behavior is unusual for ALSs and has not been observed in any other system. It stems from the progressive alignment of the magnetic domains with the global magnetic field. This can be understood more precisely from the energy dispersion of electrons subject to a rotating magnetic field, E(K) with K the wave vector, shown in Fig. 2a. The interference conditions defining the energies of the ALSs are set by the wavevectors difference $\Delta K$ between right-moving and left-moving-electrons with non-orthogonal spins. A variation of the magnetic domains induces a shift k in the wavevectors K. Near the helical gap, where the spin states are not orthogonal, it adds a term 2kL to the interference condition : $E_{ALS} \approx \pm\left(E_{ALS,0} + a\cos[\,2\Delta K(B_{ext})L]\right)$ with $\Delta\mathrm{K}(\mathrm{B}_{\mathrm{ext}}) = \Delta K(B_{ext} = 0) + 2\mathrm{k}(\mathrm{B}_{\mathrm{ext}})$. The ALSs magnetic field dependence is well accounted for by this formula, under the assumption that the spin orbit strength decays linearly as the field increases, up to a saturation field of about 1T. Such an evolution of the synthetic spin orbit energy is supported by magnetic measurements as well as micromagnetic simulations (see Extended data). The number of oscillations N sets the range of modulation of k($\mathrm{B}_{ext}$) and therefore allows us to give a lower bound for the induced spin orbit energy at zero magnetic field: $E_{so} \gtrsim \delta\, N/2$ (see Extended data). From the number of oscillations in Fig. 2b for $\mathrm{B}_{\mathrm{ext}}$>0 (N~1.5) and the extracted level spacing ($\delta$ ~1.5 meV see Extended data Fig. 2) we deduce $E_{so} \gtrsim 1.1meV$. This is of the order of the simple estimate for a linear spectrum[11,12] $E_{so} = \frac{hv_F}{2\lambda} = \delta\frac{L}{\lambda}$ with L/λ~2 (5 domains), inferred from the MFM picture in Fig. 1e. Strikingly, this spin orbit energy is larger than the ones found in InSb or InAs nanowires (respectively 0.25-1 meV and 0.015-0.135 meV[17]).

Moreover, we can reproduce the ALSs oscillations with simulations based on the scattering theory, with $\delta$ ~1.5 meV and $\Delta \sim 0.6$ meV extracted from the data, an amplitude of the stray field $B_{osc}$ of 400mT extracted from the magnetic simulations and a chemical potential close to the helical regime (see Extended data). These oscillations are robust to disorder in the magnetic texture and can be qualitatively reproduced from the spatial field evolution inferred from the MFM data of Fig. 1 (see Extended data).

In order to understand further the specificity of our device, we have realized a control device without the magnetic texture. We do not observe the presence of subgap ALSs that oscillate with the magnetic field but only the closing of the superconducting gap, and a linear splitting of a quasi-particle (QP) resonance coming from a finite density of states in the carbon nanotube (Fig. 3b). The quasiparticle resonance simply splits in $B_{ext}$ with a slope $g\mu_B$ (giving a Landé factor g~3.5), and the superconducting gap closes over $B_c = 500$mT. Under a zero or weak constant spin-orbit interaction, ALSs should display crossing oscillations with a period $\widetilde{B_{ext}}$ of the order of $\delta/g\mu_B$ (see Extended data) where $\mu_B$ is the Bohr magneton. This corresponds to $\widetilde{B_{ext}} =$ 5T for the ALSs in the control experiment, which explains why they stay pinned to the superconducting gap until it closes. Fig. 3c summarizes possible behaviors for the ALSs under a magnetic field, illustrating both experiments. In the magnetic device, the period of the oscillations with the external field is only compatible with a modulation of the induced spin-orbit interaction, through the progressive alignment of the magnetic domains. The oscillations thus point unambiguously to the non-trivial character of the observed ALSs.

The large measured value of spin-orbit interaction is an important prerequisite for driving a hybrid device into the topological regime, where zero energy Majorana modes can emerge. In all the devices experimentally investigated so far, this has only been pursued by applying a large external magnetic field, with severe constraints on network designs, Majorana mode lifetimes and coupling to superconducting quantum circuits. Contrarily, our magnetic texture is equivalent to both a finite and large spin orbit interaction and an external magnetic field: our device could host Majorana modes without any external magnetic field, thus lifting these constraints.

In our setup, superconductivity is induced from the side into the helical region, through superconducting proximity effect. Although this is a slight difference compared to other experiments, it can in principle lead to Majorana modes[17,18,19,20] (see also Extended data for a more developed discussion of this possibility). In Fig. 4a, at *zero* external field, a zero bias conductance peak emerges, simply upon tuning Gate 2 at $V_g$ > 0.5-0.6 V. It has a width of about 150 μeV as shown in Fig. 4b, and a height of about 0.05 $e^2/h$, comparable to the recent findings in semiconducting nanowires (see e.g. [7]). In Fig. 4d, we measure a large magnetoresistance of 20% for this zero bias peak, accompanied by a hysteretic behavior which is a signature of the effect of the magnetic texture. This strong dependence at small magnetic field could come from local reconfiguration of the magnetic domains, consistent with the expected spatial localization of the state corresponding to the Majorana peak, contrary to the finite energy ALSs which are not affected by a small magnetic field (see Extended data Fig. 7 and 14). Finally, Fig. 4c displays a conductance map where the zero bias peak is robustly pinned at zero energy at large external magnetic field. These features are compatible with the zero bias peak indicating the presence of a Majorana zero modes (see Extended data for more control experiments).

As a conclusion, we have demonstrated a device with a synthetic spin orbit interaction induced by a proximal ferromagnetic multilayer producing an inhomogeneous local magnetic field. This spin orbit interaction deeply modifies the superconducting correlations induced by superconducting contacts and allows us to observe a zero bias peak suggestive of a Majorana mode without any external magnetic field. By relaxing the constraint of an external magnetic field, our setup is suitable for advanced experiments that would unambiguously[22,23,24] characterize Majorana modes with the tools of cQED circuits[25,26,27,28]. The built-in 2D pattern of our magnetic textures could also be interesting for braiding schemes[29] which will require networks of Majorana modes with local and autonomous generation of topological superconductivity.

**Supplementary Information** Supplementary Information accompanies the paper on www.nature.com/nature.

**Acknowledgements.** We are indebted to B. Leridon for the SQUID measurements and to K. Bouzehouane for MFM measurements. We gratefully acknowledge J. Palomo, M. Rosticher, A. Pierret and A. Denis for technical support. L.C.C. acknowledges the support from a Foundation CFM-J.P. Aguilar grant. The devices have been made within the consortium Salle Blanche Paris Centre. This work is supported by ERC Starting Grant CIRQYS and grants from Région Ile de France and the ANR FunTheme.

**Authors contributions.** MMD setup the experiment, LC made the devices and carried out the measurements with the help of TK. LC and MMD performed the analysis of the data with inputs from TK. LC and MRD carried out the fabrication, measurement and analysis of the control experiment. MMD, JJV and LEB contributed through early experiments and to develop the nanofabrication process. TC and FM contributed to the experimental aspects. MCD developed the data acquisition software. SR and AT developed the magnetic texture process and carried out the magnetic characterization, with MMD and LC. MMD, LC, MRD, TK and AC carried out the theory for the ABS oscillations. MMD studied the tight-binding model, using a framework developed by MCD with theoretical insight from AC. TK, MMD, LC, MRD and AC co-wrote the manuscript with inputs from all the authors.

**Author Information.** Reprints and permissions information is available at www.nature.com/reprints. The authors declare no competing financial interests. Correspondence and requests for materials should be addressed to T.K. : kontos@lpa.ens.fr

**Figure 1 | Hybrid superconductor-nanotube-magnetic texture setup.**

**a.** Schematic picture of the multilayer magnetic texture with up and down domains (white and black arrows) inducing the rotating magnetic field in space ($B_{osc}$, red line) leading to the synthetic spin-orbit interaction. **b.** Zoom on the device showing the single wall carbon nanotube (in red). The bottom gate is made from a multilayer of Co/Pt. The source and drain superconducting electrodes are made out of Pd/Nb. **c.** Conductance of the device as a function of source-drain bias displaying a well-defined gap with two symmetric ALSs at energy E, shown again in the inset. The “hardness” of the gap is measured by the ratio of the conductance values marked by the star and the circle. **d.** Schematics of the Andreev-like states arising from the coupling between the nanotube and the left superconductor. The right superconductor has a residual density of states at zero bias allowing for a direct spectroscopy of the ALSs. **e.** Magnetic Force Microscope (MFM) micrograph of the device showing the magnetic texture of the bottom gate. The cut of the magnetic signal indicating field modulations (yellow and grey) along the nanotube on a scale of about 200 nm is shown at the bottom.

**Figure 2 | Oscillations of the subgap states and synthetic spin-orbit interaction.**

**a.** Left panel: Band structure arising from the synthetic spin orbit interaction with N domains. The allowed interferences in the finite length system are represented with arrows. Right panel: Schematics of how the band structure can be tuned by changing the spin orbit energy (with N' domains, the bands are shifted by $k$). **b.** Low bias conductance G map in the $V_{sd}$-$B_{ext}$ plane showing the oscillations of the ALSs (indicated by purple arrows) as a function of the magnetic field. The black lines are the fit to the theory, as described in the text and in the SI.

**Figure 3 | Control experiment and phenomenology of Andreev-like states under a magnetic field.**

**a.** Transport resonances energies with respect to $V_{sd} - B_{ext}$ in reduced units. The ALSs energy and the gap edge have been extracted from Fig. 2b (grey points). Theory fits are represented in purple and red respectively (see Extended data). **b.** Same plot as in **a** for the control experiment, without magnetic texture. A quasiparticle resonance (QP, blue) appears within the gap, with a linear blue fit (see Extended data). **c.** Table summarizing the phenomenology of the evolution of transport signatures as a function of an external magnetic field. **d.** Density of state of a 1D conductor with proximity induced superconductivity as a function of energy and $B_{ext}$ in the different scenarios of c, obtained with tight binding simulations described more in details in the Extended data Fig. 12.

**Figure 4 | Zero bias peak.**

**a.** Conductance G map in the $V_{sd}$-$V_g$ plane showing the appearance of a zero bias peak when the distant gate of the wire is tuned. **b.** Linecuts at gate voltage $V_g$=0,-1,-2,-3 V. **c.** Conductance G map at $V_g$=-3V in the $V_{sd}$-$B_{ext}$ plane showing the evolution of the zero energy peak as a function of the in-plane magnetic field. The overall background arising from the superconducting gap has been subtracted for clarity (see Extended data Fig. 3). The black lines correspond to the same fit as Fig. 2b. **d.** Low magnetic field conductance G map in the $V_{sd}$-$B_{ext}$ plane for $V_g$=-3V displaying the large magnetoresistance of the zero bias peak.

**Extended Data Figure 1 | Microwave environment of the device.** The whole device is embedded in a microwave cavity with a resonance frequency of about 7.5 GHz.

**Extended Data Figure 2 | Gate map of the Andreev Like states. a,b.** Conductance as a function of bias $V_{sd}$ and Gate 1 $V_{g1}$ showing the evolution of the Andreev Like states as a function of $V_{g1}$ for two values of Gate 2 $V_g$=0V (panel a) and -3V (panel b) such that the Zero Bias Peak is present. In panel a, the shape of the Fabry-Pérot modulations of the conductance is highlighted by the dotted black lines, and N and N+1 indicates the equilibrium charge on the dot. The level spacing $\delta$ for our quantum dot can be roughly estimated, as shown by the black arrow. In panel b, the edge of the superconducting gap and

the position of the different peaks studied are outlined. The values of Gate 1 for the different figures of the article are shown by the blue lines. **c.** Corresponding superconducting gap in log scale, at $V_g = 0V$ (bottom) and -3V (top). **d.** Comparison between the conductance measurement as a function of bias (blue line) and the corresponding fit using the Usadel equations (red dashes).

**Extended Data Figure 3 | Conductance map with background.** Raw data of the conductance as a function of bias $V_{sd}$ and the external magnetic field $B_{ext}$ at $V_g = -3V$. As shown in the map and the cuts corresponding to those of the main text, the Andreev Like resonances as well as the zero bias peak are clearly visible also in the raw data. The background originates from the peculiar shape of the density of states in the proximized Pd/Nb bilayer.

**Extended Data Figure 4 | Gap closure at high magnetic field.** Conductance map in the $V_{sd}$-$B_{ext}$ plane from 0T to 2T and back showing the gradual decrease of the superconducting gap. The map is taken at $V_g$=3V.

**Extended Data Figure 5 | Microwave power dependence of Andreev Like states and Majorana peak.** Evolution of the Andreev Like peaks as a function of the microwave power applied at the input of the cavity.

**Extended Data Figure 6 | Temperature dependence of Andreev Like states and Majorana peak. a.** Evolution of the Andreev Like peaks as a function of the

temperature from 20 mK to 1.7K for Vg=-0.6V. We interpret the higher peak for T=700mK as a small gate switch because the gate setting is close to the transition at which the zero bias peak emerges. Such a switch is absent in panel B which is for a gate setting further into the gate region where the zero bias peak appears. **b.** Evolution of the Andreev Like peaks as a function of the temperature from 50 mK to 800mK for Vg=2.5V.

**Extended Data Figure 7 | Hysteresis of the Majorana peak. a-e.** Conductance map $B_{ext}$-$V_{sd}$ at small magnetic field for different gate voltages. The presence of the vertical stripe corresponding to the magnetoresistance is correlated to the emergence of the Majorana peak. **f.** Difference in conductance between upward and downward field sweeps at zero bias showing the gate voltage dependence of the hysteresis.

**Extended Data Figure 8 | Characterization of the control device. a.** Conductance map $V_g$-$V_{sd}$ of the control device for different external magnetic fields, $B_{ext}$=0T and $B_{ext}$ =1.7T. **b.** Conductance map $B_{ext}$ -$V_{sd}$ at a gate voltage corresponding to the dashed line in a.

**Extended Data Figure 9 | Magnetic characterization of CoPt multilayer. a.** SQUID measurement at 4K of a 5mm x 5mm chip covered with the CoPt multilayer (as sketched above). The magnetization saturates at about 1.5T.

Inset: Zoom on the SQUID measurement showing the opening of a hysteresis at about +/-50mT, of width 20 mT. **b.** SQUID measurement at 4K of a 5mm x 5mm chip covered with an array of the nanoscale Co/Pt stripes (such as the ones used in the transport experiment, see the layout above). The magnetization saturates only at about 2.5T. Inset: Zoom on the SQUID measurement showing the opening of a hysteresis at about +/- 100 mT, of width 20 mT. **c.** Magnetization texture for zero effective anisotropy and 15% roughness. The white and black pixels correspond respectively to up and down magnetization, the colored pixels represent the in-plane magnetization, colored according to a color wheel to represent their different orientation (red correspond to the applied field direction, see the arrow). The bottom image, numbered 1, corresponds to a virgin demagnetized state; the following are successive images from 0.25 T to 1 T (numbered 2 to 4). Images are 768 nm x 2304 nm, similar to the experimental Co/Pt gate dimensions. **d.** Calculated hysteresis loops for in-plane magnetic field, for two anisotropy hypotheses and two magnitude of roughness. **e.** Cuts of the magnetic field along the dashed line in c, obtained from the same magnetic simulation. The $B_x$ (resp. $B_z$) field is represented in blue (resp. red), at a height x=10nm.

**Extended Data Figure 10 | MFM image of the same Co/Pt structure in another device.** Topography and phase contrast of a Co/Pt structure identical to the one of the magnetic texture device, evaporated in a trench and imaged afterwards, without any additional process as opposed to the main device. Here the MFM signal is much more regular.

**Extended Data Figure 11 | Analysis of the Andreev Like states oscillations.** **a.** Schematic of the scattering representation of the device. The magnetic texture is modelled by a field helix over a length $L_1$, and is surrounded by two short segments of length $L_2$ with a uniform magnetic field. It is connected to a superconductor on one side. **b.** Energy levels of the system as given by equation $(2)$, as a function of energy and number of field oscillations, which is directly linked to the spin-orbit energy $E_{so}$. With parameters coherent with our experiments, we are able to reproduce several oscillations of the ALSs emergent in this device. **c.** Convolution of the density of state of a device with 2 pairs of ALSs (with energies corresponding to $0.3\ \Delta$ et $0.7\ \Delta$, a spacing that can be obtained with a slightly higher $B_{osc}$) and a degraded superconducting density of state (as schematized in Fig. 1). One pair of ALSs is hidden in the slope of the conductance as a function of applied bias.

**Extended Data Figure 12 | Oscillations of the Andreev like states in the various situations mentionned in the table of the main text.** The colorscale map displays the density of states (DOS) as a function of the energy $eV/\Delta$ . Panel a (resp b) displays the evolution of Andreev Bound states as a function of a homogeneous magnetic field $B_{ext}$ without (resp with) a spin-orbit interaction in the chain. Panel v displays the evolution of ALS with respect to the number of oscillation of  a cycloidal field in the normal part. The density of states in c shows non-crossing oscillations as a function of the number of oscillations.

**Extended Data Figure 13 | Effect of disorder in the oscillating magnetic field. a.** Cut of the MFM image along the CNT as represented in Fig. 1 of the Main Text (in blue), and reconstructed signal containing the N=40 top-most frequencies of the discrete Fourier transform of the raw data (orange). **b.** Modulus of the Fourier coefficients of the MFM signal of a. A given discrete frequency can be thought of as the number of up and down domains in the magnetic texture. To mimic the effect of the external magnetic field, we shift the coefficients' frequency as shown in equation (3). For example, the coefficient corresponding to 5 oscillations at the beginning will, at $\delta f = 2$, correspond to 7 oscillations. **c.** Density of state of a one-dimensional wire as a function of the energy showing Andreev Like States (ALS). The oscillating field is given by the orange curve of panel a and evolves under a shift of it Fourier coefficients. The x axis corresponds to shifting frequencies. **d.** The oscillating magnetic field is now taken from the magnetic simulations, in all three directions of space. **e.** Similar plot, with only one coefficient in the discrete Fourier transform. **f.** Oscillating magnetic field in all three directions of space, extracted from the magnetic simulations, along the dashed line in Extended Data Fig. 7d. It is given for different external magnetic fields, at a height x=0nm.

**Extended Data Figure 14 | Emergence of Majorana excitations in our experimental setup. a.** Density of state at the first site of the chain (site 0), computed using a discretized tight-binding Hamiltonian, showing the emergence of a zero-bias peak for a large number of up and down domains. **b-c.** shows the singlet and triplet pairing amplitude for the same parameters. Singlet and triplet

pairing amplitude are defined in the text of this supplementary. The numerical simulation is realized by considering a normal part with a cycloidal field, as illustrated in d for the component along the axis of the chain (z-axis). The field is not perfectly sinusoidal because of high ratio between the number of oscillations (7) and the number of sites (16) for the cycloid. A superconductor is connected to the chain from sites 15 to 20. Two local reconfigurations of the field are shown in red and blue. e, f and g display the spatial dependence of the density of states for the three different configurations of the field (black, red and blue). The dotted line shows the separation between the normal part and the superconducting part.

## METHODS

**Fabrication of the devices and measurement techniques.** A 150nm thick Nb film is first evaporated on a high resistivity Si substrate at rate of 1nm/s and a pressure of $10^{-9}$ mbar. A microwave cavity (see next section) is made subsequently using photolithography combined with reactive ion etching ($SF_6$ process). An array of bottom gates is then made with two e-beam lithography steps in a $100\ \mu m$ square opening of the cavity ground plane near the central conductor. First, we etch 750nm x 36 $\mu m$ trenches of 60nm depth with reactive ion etching ($CHF_3$ process). Second, we deposit inside the trenches a Ta(4nm)/Pt(6nm)/$[Co(2nm)/Pt(1.2nm)]_{10}$/Pt(4.8nm)/Al(4nm) multilayer bottom gate, 100 nm narrower. This magnetic stack has been chosen to promote spontaneously magnetic textures, consisting of magnetic stripes with up and down magnetization direction with a narrow period. The Pt/Co interfaces induce a perpendicular anisotropy energy that partially compensates for the shape anisotropy which would induce an in-plane magnetization. The Co thickness as well as the number of repetitions have been chosen to increase the dipolar energy, the driving force of the stripes formation, and to maximize the stray field above the sample to about +/-400 mT. All the layers in the sample are strongly coupled through their magnetic stray field and belong to a single and continuous magnetic texture. Carbon nanotubes are grown with Chemical Vapor Deposition technique (CVD) at about 900°C using a methane process on a separate quartz substrate and stamped above the bottom gates[20]. The quartz substrate was previously processed so that a few pillars of height 4 $\mu m$ and surface 10 $\mu m$ x 5 $\mu m$ are aligned with the cavity openings and come in contact when stamping (see the stamping marks in Fig. 1a). The nanotubes are then localized and those which correctly lie on a bottom gate are contacted with Pd(4nm)/Nb(40nm). The Nb layer is evaporated at a rate of 1 nm/s and pressure of $10^{-9}$ mbar while the substrate is cooled down at 0°C. During this last e-beam lithography and evaporation step, gate electrodes are also patterned in order to couple capacitively the bottom gate to a DC gate voltage $V_{g1}$ (named Gate 1 in the following) and to the AC potential of the central conductor of the cavity. An additional gate, Gate 2 is capacitively coupled to the central core of the cavity, and a voltage $V_g$ can be applied.

The DC measurements are carried out using standard lock-in detection techniques with a modulation frequency of 137 Hz and an amplitude of 10 $\mu V$. The base temperature of the experiment is 18 mK. An external magnetic field can be applied along the direction of the tube.

A control device was fabricated with a slightly different fabrication technique, and the Co/Pt gate was replaced by an Ti(5nm)/Al/AlOx(100nm) bottom gate. The chemical potential is tuned through an additional gate that forms a fork around the bottom gate, noted $V_g$ in the characterization. In a similar fashion, the carbon nanotube was connected to two Nb(150nm)/Pd(4nm) superconducting electrodes. The DC measurement are carried at a modulation frequency of 77,77 Hz and an amplitude of 15 $\mu V$ since the signal is smaller.

**Large scale device layout and microwave environment.** The large scale device layout and microwave environment is shown in Extended data Fig. 1. The whole device is embedded into a microwave cavity which has a fundamental resonance frequency of about 7.5 GHz. For the device presented in the main text, this particular mode was not coupled to the device but other modes of the electromagnetic environment were coupled. We use here these modes to couple our device with a distant gate Gate2 (with gate voltage $V_g$) which is the one used in the main text. The device is also coupled directly via the bottom gate to Gate 1 with gate voltage $V_{g1}$.

Although the specific resonant mode of the cavity was not coupled to our device, the coplanar waveguide resonator could be used to convey a microwave signal in the GHz range to study its dynamical response (see below).

**Usadel equations for accounting for the conductance at high bias.** We present in this section the theory which allows us to account well for the shape of the full conductance curves based on the quasiclassical theory of superconductivity in the diffusive regime (Usadel equations). We use a superconducting bilayer of Nb(40nm)/Pd(4nm) to make a superconducting contact on the nanotube. The density of states in these bilayers are in general non BCS because of interface resistance between the superconducting slab and the normal slab and inverse proximity effect. In addition, in our case, the dipolar helicoïdal-like field can induce a pair-breaking which can be taken into account via an Abrikosov-Gorkov general term. The Usadel equations read:

$$\frac{\hbar D}{2}\partial_z^2\vartheta(z) - (-iE + \gamma)sin\vartheta(z) - 2\Gamma_{AG}sin\vartheta(z)cos\vartheta(z) + \Delta(z)cos\vartheta(z) = 0$$

where $E$ is the energy, $D$ is the diffusion constant, $\Gamma_{AG}$is the Abrikosov-Gorkov pair-breaking parameter, $\gamma$ is the "Dynes" parameter and $\Delta(z)$ is the gap function. The pairing angle $\vartheta(z)$ is related to the normal and anomalous Green's functions, G and F respectively, via the relation: $G(z) = cos\vartheta(z)$ and $F(z) = sin\vartheta(z)$. The density of states is $N(z, E) = Re(cos\vartheta(z))$.

An important energy scale controlling the physics of proximity effect in the bilayer is the Thouless energy $E_{Th} = \frac{\hbar D}{d^2}$, where d is the thickness of the normal (Pd) slab. In our case, this energy is much larger than the superconducting gap $\Delta$ of the homogeneous superconductor, and the interface is not too opaque. Neglecting self-consistency, one may approximate the pairing angle by the homogeneous solution which obeys the following implicit equation:

$$tan\vartheta_0 = \frac{\Delta}{-iE + \gamma + 2\Gamma_{AG}cos\vartheta_0}$$

The above equation may be solved numerically and the gap appearing in it has the meaning of an effective gap in the normal slab. The density of states in each Nb/Pd bilayer contacting our nanotube is therefore $N(\Delta, \gamma, \Gamma_{AG}, E) = Re(cos\vartheta_0)$.

In order to compute the current I flowing through our device and the corresponding conductance G=dI/dVsd, one may use the above density of states. As explained in the main text, one of the two contact is a tunnel probe and the nanotube is only in good contact with the other. One can therefore approximate the density of states of the latter by that of the Nb/Pd bilayer which induces the superconducting correlation in it and a sum of two lorentzians describing the Andreev Like states:

$$N_{NT}(E) \approx N(\Delta, \gamma, \Gamma_{AG}, E) + \beta \sum_{i=+/-} \frac{\eta}{(E - i \times E_{SGS})^2 + (\eta/2)^2}$$

This equation is an approximation since it neglects the transfer of spectral weight from the density of states of the slab to the Andreev states and is only valid as long as $\beta << 1$. The tunnel current can be expressed using the usual tunnel spectroscopy formula:

$$I = G_N \int_{-\infty}^{+\infty} dE\, N_{NT}(E) N\big(\Delta, \gamma_2, \Gamma_{AG,2}, E + eV_{sd}\big)\{f(E) - f(E + eV_{sd})\}$$

where $f(E)$ is the Fermi function. The above formula is the one used to fit the cut in Fig. 2a. We allow the Abrikosov-Gorkov $\Gamma_{AG}$ and the Dynes parameter $\gamma$ to be different for the left and the right tunnel contact.

The fit presented in Extended data Fig. 2d was obtained with $E_{SGS} = 210$ μ$eV$, $\eta = 79$ μ$eV, \Delta = 716$ μ$eV$, $\Gamma_{AG} = 51$ μ$eV$, $\Gamma_{AG,2} = 318$ μ$eV$, $\gamma_1 = 16$μ$eV$, $\gamma_2 = 82$ μ$eV$, $\frac{4\beta}{\eta} = 0.74$, T=150mK and $G_N = 0.61\ \frac{e^2}{h}$. This leads to two density of states $N_{NT}$ (contact 1) and $N$ (contact 2), plotted in Fig. 2b. It also show that we can safely remove that background from the bare curves shown in Extended data Fig. 3 to plot those of Fig. 4 in the main text.

Importantly, this fit also shows that the very low value of conductance associated to the two Andreev Like states does not imply low spectral weight. This is best seen on Fig. 2b where the peak height of the two Andreev Like peaks amounts to about 0.74 of the normal state density of states in the nanotube. Similarly, the zero bias peak has a spectral weight much larger than its height in conductance. From the comparison with the height of the Andreev Like states which are roughly twice as large, we can estimate that the actual peak height of the zero bias peak is about 0.35 of the normal state density of states in the nanotube.

**Additional data on the magnetic texture device: Gate map of Andreev Like states.** We present the gate map of the Andreev Like states when Gate 2 is kept at 0V and Gate 1 (which is directly coupled to the bottom gate as shown in Extended data Fig. 1) is swept from 0.2V to 1V in Extended data Fig. 2. The Andreev Like states remain visible essentially in all the map. A parity crossing is observed at $V_{g1}$~0.7V. Importantly, the high bias conductance, close to $e^2/h$ displays only weak features. In particular, no Coulomb bockade diamond is observed which signals that our experiment is in the Fabry-Pérot regime. From the smooth chessboard pattern, one can extract an estimate of the level spacing δ~1.5 meV.

**Conductance maps with background and gap closure at high magnetic field.** We present in this section two conductance maps corresponding to those of the main text. In Extended data Fig. 3, we present the raw data corresponding to Fig. 4a. In this map, one can see that there is a strong depression of the conductance as a consequence of the superconducting gap. After fitting the above curves with the theory presented above, one can extract the contribution arising only from the Andreev Like states which allows one to observe more clearly the magnetic field dependence of these states. Nevertheless, as one can see in Extended data Fig. 3, all the features presented in the main text are visible in the raw data.

Finally, it is interesting to study the magnetic field map of the conductance up to large fields where the superconducting gap of the electrodes starts to weaken substantially. In Extended data Fig. 4, we present such a map where the magnetic field is swept from 0T to 2T and back to 0T. As expected, we observe a gradual “square root like” decrease of the gap edge.

**Control experiments: Microwave power and temperature dependence of Andreev Like states and Majorana peak.** In this section, we present two control experiments. In the first, we apply a large microwave power to the input port of the microwave cavity in order to test whether the zero bias peak may arise from a weak Josephson effect. In the adiabatic limit where the frequency of the applied tone to the cavity $f_{rf} = 5{,}6$ GHz is much smaller than the relevant relaxation rates of the states in the device, the conductance $G$ is modulated by the cavity photons as:

$$G(t) = G\big(V_{sd} + V_{AC}\cos(2\pi f_{rf} t)\big)$$

The conductance can be fit by 3 lorentzians centered around each of the peak energies as shown in the section devoted to the finite bias conductance. The phenomenology of the above equation is simply a splitting of each conductance peaks if $V_{AC}$ bias larger than their width. As shown in Extended data Fig. 5, the two finite energy Andreev Like states as well as the central peak split at the same power and in the same way showing that they all correspond to electronic states characterized by a lorentzian like spectral density. In particular, these measurements are not consistent with the zero bias peak being a well-developed Josephson supercurrent branch which would display Shapiro steps. The case of a weak Josephson branch which does not display Shapiro steps would be very quickly washed out by temperature (at the temperature scale given by the Josephson energy, ie a supercurrent of 1nA corresponds to 140mK ) and is not consistent with the temperature dependence of the zero bias peak which is described below.

Finally, we present in Extended data Fig. 6 the temperature dependence of our measurements which is fully consistent with a gradual filling of the gap which starts to be effective only at about 1K. In particular, as one can see in panel A of Extended data Fig. 6, we observe that the zero bias peak and the Andreev Like states disappear at the same temperature (about 1K). Therefore, we can exclude a thermal occupation origin for the zero bias peak that would be indicated by a continuous increase of the zero bias peak as a function of temperature.

**Magneto-resistance and hysteresis at different gate voltages $V_g$.** Extended data Fig. 7 displays a panel of the conductance maps $V_{sd} - \mathrm{B_{ext}}$ for different gate voltages between 0 and -3V. We observe that the ALS are insensitive to small magnetic fields, whereas the zero bias peak and the background shows a magneto-resistance provided the zero bias peak is present. These two different behaviors are also observed in the Gate 2 dependence that leaves the ALS unchanged. However we cannot match this magnetoresistance with a shift in gate voltage. We also present the emergence of the hysteresis of the zero bias peak with the gate voltage which is directly linked to the emergence of the peak.

**Characterization of the control device.** Extended data Fig. 11 shows additional characterization of the control device presented in Fig. 3 of the main text.

The control device displays clear Coulomb diamond as shown in Extended data Fig. 8a, at 0T and 1.7T. Although the two contact electrodes are made of Nb, the conductance maps can be interpreted as transport signatures through a S/QD/N system, as in the case with a magnetic texture device.

At 0T, we see a clear superconducting gap in the transport features. Residual density of state in the superconducting contact gives rise to a weak quasiparticle peak below the superconducting gap. Extended data Fig. 11b shows the evolution of these transport peaks as a function of an in-plane external magnetic field, from which the peaks positions of Fig. 3b are extracted. The superconducting critical field for the control device is lower than the one of the magnetic texture device due to the change in the Nb electrode thickness (150nm instead of 40nm).

We use a constant interaction model to obtain the stability diagram equations for a S/QD/N system, used in the fit of Fig. 3a. The quasiparticle peaks positions is given by $eV_{sd} = \frac{\epsilon(B)}{\alpha}$ and $eV_{sd} = -\frac{\epsilon(B)}{1-\alpha}$, where α is the contact asymmetry and $\epsilon(B) = (\epsilon_0 - g\,\mu_B\,B)$ is the chemical potential of the dot (for a spin down electron, in agreement with the diamond shifting to the left with magnetic field as seen in Extended data Fig. 8a).

The superconducting gap peaks positions are given by $eV_{sd} = (\Delta(B) - \epsilon(B))/(1 - \alpha_2)$ (at positive bias) and $eV_{sd} = -\frac{\Delta(B)+\epsilon(B)}{1-\alpha}$ (at negative bias) where $\Delta(B) = \Delta\sqrt{1 - \left(\frac{B}{B_c}\right)^2}$. Here we used different contact asymmetries between positive and negative bias to fit the data, as one can see in Extended data Fig. 8a that the slope does change.

One can note that since the diamonds shifts in energy, the lower gap evolution should be piecewise-defined. However the critical field is reached before this is needed.

The fit values are the following: Δ = 0.68 meV; $B_c$ = 0.6 T; $\epsilon_0$ = -0.02 meV; g = 3.8; α= 0.31; $\alpha_2$ = 0.47.

**Magnetic characterization of the CoPt multilayer.** We present in Extended data Fig. 9 magnetic characterization of our CoPt multilayers. Our multilayers are characterized by SQUID magnetometry with an in- plane magnetic field, both on a plain substrate deposition (panel A) and on a chip covered of 650 nm x 30 μm stripes, processed exactly as the ones used for our bottom gates and thus expected to have the same amount of disorder (panel B). After a sharp increase for low magnetic field, the magnetization displays a slow saturation. Although the hysteresis span stays about 20 mT, the nanostructuration increases the magnetic field range for the hysteresis from 100mT to 200mT (see insets) and the saturation field from 1.5T to 2.5 T. The magnetization values at saturation differ between the two measurements due the uncertainties in the value for the volume of Co. Whatever the orientation of the applied magnetic field or the magnetic history, a low remanence is found indicating that the sample spontaneously demagnetizes. The presence of magnetic textures at zero field is doubtless in the demagnetized state, as further confirmed by magnetic force microscopy (see Fig. 1c). When approaching the saturation, the nature of the magnetic state has to be understood.

To further understand the magnetization processes in the sample, we have performed micromagnetic simulations, using the MuMax3 code[30]. As the magnetic parameters are not exactly known, the purpose is not to reproduce exactly the sample under study, but to understand qualitatively the processes, and in particular explain the field strength needed to saturate magnetization in the samples. The saturation magnetization used ($1.2\text{x}10^6$ A/m) has been extracted from SQUID data (panel A) and the exchange (10 pJ/m²) is the one of bulk cobalt. As the magnetic anisotropy could not be measured, we explored two hypotheses: zero effective magnetic anisotropy [exact compensation between shape ($-\frac{1}{2}\mu_0 M_s^2 = 0.905 \times 10^6 \mathrm{J}/m^3$) and interface induced anisotropies ($\frac{K_s}{\mathrm{t}} = 0.905 \times 10^6 \mathrm{J}/$ $m^3$ with t the Co layer thickness)] and a small negative effective anisotropy ($\frac{K_s}{\mathrm{t}} = 0.7 \times 10^6 \mathrm{J}/m^3$) thus favoring in-plane magnetization orientation for homogeneous magnetized states with an effective anisotropy $K_{eff} = \frac{K_s}{\mathrm{t}} - \frac{1}{2}\mu_0 M_s^2 = -205 \times 10^6 \mathrm{J}/m^3$. These hypotheses are in agreement with litterature for the interface anisotropy at the Co/Pt interface[31]. For both cases, whatever the magnetic history is, the ground state corresponds to a demagnetized state with periodic stripes[32] (~ 100 nm period). Upon applying a magnetic field, the stripes first reorient progressively with a propagation vector orthogonal to the magnetic field to minimize the Zeeman energy in the domain walls (the domain walls being Bloch-type, their magnetization lies in the wall plane). For larger fields, the magnetization in the domains progressively rotates toward the applied magnetic field direction and saturates at about 1T. Note that textures are still observed up to the saturation. The saturation field value is to be compared to the anisotropy field $\mu_0 H_K = 2\frac{K_{eff}}{M_s}$, which is close to zero. In usual magnetic system with a low demagnetizing strength (low magnetization systems or low thickness), this would cause a saturation at field values close to $H_K$. Here, the large magnetization and the ten repetitions, both favoring stripe phases, make the saturation much more difficult and therefore result in a saturation field which scales with the magnetization value. Comparing the calculated and the experimental loops we note that the saturation is much slower in the sample than in the calculation. We attribute this effect to the disorder in the sample. Indeed, due to the fabrication process the quality of the substrate could not be optimized, which results in a significant roughness. While typical roughness in good magnetic samples is about 2-5%, here larger values could be expected. We have calculated the magnetic loop with increasing roughness up to 15% thickness variation, using the roughness model successfully developed in previous studies[33,34]. While the loops are not much changed at small field, we note that due to the disorder, saturation occurs at much larger magnetic fields and the images show that magnetic textures may survive up to 2T. This validates our assumption of a smooth decrease of the synthetic spin orbit interaction used to account quantitatively for the transport data.

Additional MFM measurement under an external magnetic field were obtained, between 0 and 1.2T. Since the MFM signal is proportional to $\frac{d^2B_z}{dz^2}$ , it can only give a qualitative image of the magnetic texture and as such these measures are not reproduced here. However we notice that the magnetic domains persist up to 800 mT, in rough agreement with the saturation field $B_S$ of the observed transport oscillations (introduced below). We also noted that for the nanostructured SQUID device, the domains disappeared at lower magnetic field, illustrating the influence of disorder in our system. As a consequence, the SQUID data, even for the nanostructured device, is not perfectly representative of our device.

**MFM characterization of another multilayered structure.** We present in Extended data Fig. 10 the AFM (height) and MFM (phase contrast) image of a trenched Co/Pt multilayered structure identical to the one of the magnetic texture device, in a more controlled environment since there was no additional processing after the deposition.

**Analytical theory of the oscillations of Andreev Like states with magnetic field.** We present in this section the non-interacting theory accounting for the oscillations of the Andreev Like states as a function of the magnetic field. The hamiltonian of the system in the normal state can be written as:

$$\widehat{H} = -\left(\frac{\hbar^2 \partial_z^2}{2m} - \mu(z)\right) + \frac{1}{2} g\mu_B \overrightarrow{B_{osc}}(z).\vec{\sigma} \qquad (1)$$

where $\vec{\sigma}$ is the spin operator of electrons and $\vec{B}(z)$ is the rotating magnetic field acting on the electron spin. The above hamiltonian can be “integrated” in order to calculate the transfer matrix $\mathcal{T}$ of a section of 1D system of length L subject to the rotating field. In the case of a regular cycloidal field which rotates at a speed : $\frac{\partial\theta}{\partial z} = \alpha$, we can write the transfer matrix $\mathcal{T}$ as:

$$\mathcal{T} = exp\{i(\widehat{K} + \alpha\hat{\mathcal{A}})L\}$$

The matrices $\widehat{K}$ and $\hat{\mathcal{A}}$ take the following expressions:

$$\widehat{K} = \begin{bmatrix} k_\uparrow & 0 & 0 & 0 \\ 0 & k_\downarrow & 0 & 0 \\ 0 & 0 & -k_\uparrow & 0 \\ 0 & 0 & 0 & -k_\downarrow \end{bmatrix} \quad \text{and} \quad \hat{\mathcal{A}} = \frac{1}{4\sqrt{k_\uparrow k_\downarrow}} \begin{bmatrix} 0 & k_\uparrow + k_\downarrow & 0 & k_\uparrow - k_\downarrow \\ -(k_\uparrow + k_\downarrow) & 0 & k_\uparrow - k_\downarrow & 0 \\ 0 & k_\uparrow - k_\downarrow & 0 & k_\uparrow + k_\downarrow \\ k_\uparrow - k_\downarrow & 0 & -(k_\uparrow + k_\downarrow) & 0 \end{bmatrix}$$

with $k_\sigma = \sqrt{\frac{2m}{\hbar^2}(E + \mu + \sigma B_{osc})}$, $\mu$ being the chemical potential of the wire, $E$ the energy of the electron and $B_{osc}$ the amplitude of the oscillatory magnetic field. Interestingly, the eigenmodes of the matrix $\widehat{K} + \alpha\hat{\mathcal{A}}$ allow us to define “energy bands” even in the finite size system. We find two eigenmodes $k_\pm$ which allows us to find the two bands:

$$E_\pm = -\mu + \frac{E_{SO}}{4} + \frac{\hbar^2 k_\pm^2}{2m} \pm \sqrt{\frac{\hbar^2 k_\pm^2}{2m} E_{SO} + \left(\frac{g\mu_B B_{osc}}{2}\right)^2}$$

with $E_{SO} = \frac{\hbar^2\alpha^2}{2m}$. This is exactly the same bands than with a Rashba spin orbit interaction and an external magnetic field suitable for the emergence of Majorana zero modes. The corresponding energy bands and how they vary as a function of the spin orbit energy are depicted in Fig. 2 of the main text.

In order to refine the model and take into account the two end sections with homogeneous stray field as measured from the MFM, we allow two sections before and after the oscillating field region to be partially polarized by a magnetic field. The full transfer matrix of the 1D system depicted in Extended data Fig. 5 now reads:

$$\mathcal{T}_{tot} = R(0,\mu_L,E)^{-1}R(h_{pol},\mu_L,E)exp\{i\widehat{K}_L L_L\}R(h_{pol},\mu_L,E)^{-1}R(h,\mu,E)exp\{i(\widehat{K} + \alpha\hat{\mathcal{A}})L\}R(h,\mu,E)^{-1}R(-h_{pol},\mu_R,E)exp\{i\widehat{K}_R L_R\}R(-h_{pol},\mu_R,E)^{-1}R(0,\mu_R,E)$$

where $R(h,\mu,E)$ is the transfer matrix of each interface represented in Extended data Fig. 5.

$$R(h,\mu,E) = \begin{bmatrix} \frac{1}{\sqrt{k_\uparrow}} & 0 & \frac{1}{\sqrt{k_\uparrow}} & 0 \\ 0 & \frac{1}{\sqrt{k_\downarrow}} & 0 & \frac{1}{\sqrt{k_\downarrow}} \\ \frac{1}{\sqrt{k_\uparrow}} & 0 & -\frac{1}{\sqrt{k_\uparrow}} & 0 \\ 0 & \frac{1}{\sqrt{k_\downarrow}} & 0 & -\frac{1}{\sqrt{k_\downarrow}} \end{bmatrix}$$

In this modeling, we assume that we have two sections arount the cycloidal region in which the electrons propagate under a homogeneous magnetic field. In accordance with the magnetic simulations which shows two opposite magnetic charges at the end of the cycloïdal section (due to the contribution of the inter-domains regions), we take these fields to be of the same magnitude but opposite.

The transfer matrix $\mathcal{T}_{tot}$ allows us to determine the scattering matrix $S_R(E)$ of the 1D section in the absence of superconductivity and terminated by a wall (see Extended data

Fig. 5). In the presence of a superconducting reservoir, the Andreev Like states energies $E_{ALS}$ may then be found using the following identity[35] stemming from the secular equation of the system:

$$Det\{\mathbf{1} - \gamma^2 S_R(E)\hat{\sigma}_y S_R(-E)\hat{\sigma}_y\} = 0 \qquad (2)$$

where $\gamma = e^{-i\, arccos(E/\Delta)}$ is the Andreev reflection amplitude and $\hat{\sigma}_y$ is the y axis Pauli matrix.

Extended data Fig. 11 shows a colorscale map of the Andreev Like states energies obtained from the secular equation (2) as a function of $\alpha$. $\alpha/2\pi$ corresponds to the number of field oscillations, related to the number of domains in the experiment. We are able to reproduce the observed oscillations of the Andreev Like states with reasonable physical parameters using the model described above and depicted in Extended data Fig. 11. The parameters used are the following (assuming a Landé factor of g=3.5, a similar value as the one of the control device):

$$B_{osc} = 0.47\,T, \qquad B_{pol} = 1.5\,B_{osc}, \qquad \Delta = 600\,\mu eV,$$
$$\delta = \frac{\hbar^2}{2mL^2} = 0.6\,meV\,,\ L_L = L_R = L\,,\ \mu = 0.4\,\Delta, \qquad \mu_L = \mu_R = 0.5\,\Delta.$$

**Parameters used for the fit for the Fig. 3a.** The oscillations in Extended data Fig. 11 are well fitted by a simple sinusoidal function. As a consequence, we fit our oscillation data with the following heuristic formula derived from the fact that they stem from interference effect and considering that the spin-orbit decreases linearly with the applied magnetic field up to a saturation value $B_s$:

$$eV_{sd} = \pm\left(E_{ALS,0} + a\cos[\,2\Delta k(B)L]\right) = \pm\left(E_0 + \epsilon\cos\left(\frac{2\pi B}{\tilde{B}_{ext}} + \phi_0\right)\right) \text{ for } B < B_s$$

The fitting parameters are $B_s$= 0.9T; $\tilde{B}_{ext}$ = 0.56 T; $\epsilon$ = 0.018 meV; $E_0$= 0.195 meV ; $\phi_0$= 0.04 .

We also include in the fit the closing of the superconducting gap:

$$eV_{sd} = \Delta\sqrt{1 - \left(\frac{B}{B_c}\right)^2} \text{ with } \Delta = 0.45 \text{ meV and } B_c = 1.92 \text{ T.}$$

**Large doping limit (Zeeman induced oscillations).** In the large doping regime which allows us to linearize the dispersion relation: $k_\sigma^e \approx k_0 + \frac{E}{\hbar v_F} - \sigma\frac{g\mu_B B_{osc}}{2\hbar v_F}$ for the electrons and $k_\sigma^h \approx k_0 - \frac{E}{\hbar v_F} + \sigma\frac{g\mu_B B_{osc}}{2\hbar v_F}$, where $k_0$ is the Fermi wave vector. In this limiting case, equation (4) becomes, for each spin $\sigma$:

$$1 = \gamma^2 e^{2i\left(k_\sigma^e - k_\sigma^h\right)L}$$

Specifically, this equation yields the following implicit equation:

$$E_{ALS} = \pm\Delta\, cos\left\{2\pi\left(\frac{E_{ALS}}{\delta} + \sigma\frac{1}{2}\frac{g\mu_B B_{osc}}{\delta}\right)\right\}$$

In the limit of large level spacing $\delta \gg \Delta$, the above equation simply becomes:

$$E_{ALS} = \pm\Delta\, cos\left\{\pi\frac{g\mu_B B_{osc}}{\delta}\right\}$$

In case the Andreev Like states are only subject to an external magnetic field $B_{ext}$ (pure Zeeman effect), their evolution is obtained by making the substitution $B_{ext} = B_{osc}$. The Andreev Like states oscillate as a function of the external magnetic field and cross at zero energy when $g\mu_B B_{ext} = \frac{\delta}{2} + n\delta, n\epsilon\mathbb{Z}$. In our case, one oscillation would require a field of 5T, an order of magnitude larger than the observed period of 600 mT.

**Numerical study of the synthetic spin-orbit: oscillations of Andreev-like states and emergence of Majorana zero energy modes.** In order to investigate numerically the evolution of Andreev-like states with respect to different scenarii, we consider the discretized version of the hamiltonian (1) :

$$H = \sum_{n\in[1,N_1]} d_n^\dagger(-\mu\hat{\sigma}_0 + B_{osc,z}(n)\hat{\sigma}_z + B_{osc,x}(\mathrm{n})\hat{\sigma}_x + B_{osc,y}(\mathrm{n})\hat{\sigma}_y)\, d_n - t(d_n^\dagger d_{n+1} + d_n^\dagger d_{n-1})\hat{\sigma}_0 + \sum_{n\in[N_2,N_{tot}],k} t_{kn} d_n^\dagger c_k \hat{\sigma}_0 + h.c. + H_S$$

where $d_n^\dagger = \{d_{n\uparrow}^\dagger, d_{n\downarrow}^\dagger\}$ and $d_{n\sigma}^\dagger$ is the creation operator of an electron at site n and $c_k^\dagger = \{c_{k\uparrow}^\dagger, c_{k\downarrow}^\dagger\}$ , the creaction operator of an electron in the superconductor with momentum k, and $H_S$ the hamiltonian of the superconductor. The chain with sites label by n is along the z-axis. The superconductor is coupled to the chain only between sites $N_2$ and $N_{tot}$. For the sake of simplicity, we take $t_{kn} = t_S$. A normal part between sites 1 and $N_1$ is subject to a magnetic field $B_{osc,x,y,z}(n)$. We calculate from this hamiltonian the retarded Green's function in the Nambu x Spin space at each site which allows us to obtain the conductance through the system through a Meir-Wingreen formula and the pairing function through the anomalous propagators (see for example[33]).

Extended data Fig. 12 displays three different scenari :

A. The oscillating field is set to zero, $B_{osc}(n) = 0$ and we look at the evolution of the density of states at the first site as a function of an homogeneous field $B_{ext}$ applied in the whole chain. The Andreev bound states display crossings at zero energy, with a period set by the energy level spacing $\delta$. The level spacing is obtained by looking at the conductance with respect to the chemical potential $\mu$, in the absence of the superconductor, $N_1 = N_{tot}$.

We show a simulation for $N_{tot}$ = 60, $N_1$ = 30, $N_2$ = 30, t=100, $\Delta = 1, t_s = 100, \Gamma_N = 0, \gamma_n = 0.1, \mu = 0$.

B. In scenario B, we look at the evolution of the Andreev Like states with respect to an external magnetic field but with a finite spin-orbit energy in the chain, modeled in the discrete Hamiltonian by an additional term : $\sum_{n\in[1,N_{tot}]} \Lambda d_n^\dagger \hat{\sigma}_y d_{n+1} + h.c.$ The Andreev Like States display anti-crossing at small magnetic field on a period which is bigger than the energy level spacing $\delta$. We show a simulation for $N_{tot}$ = 60, $N_1$ = 40, $N_2$ = 20, t=100, $\Lambda = 20$, $\Delta = 1, t_s = 1, \Gamma_N = 0, \gamma_n = 0.1, \mu = -0.99 * 2t$.

C. We then consider the scenario where the external magnetic field shifts the number of oscillations of a cycloidal field in the normal part: $B_{osc,x}(n) = B_{osc} * \cos(2\pi n\alpha)$ and $B_{osc,z}(n) = B_{osc} * \sin(2\pi n\alpha)$. We take into account the stray field out of the magnetic texture by implementing a homogeneous field in the superconducting part with amplitude 0.5 $B_{osc}$. The ALS show similar oscillations as observed in the experiment for which an external magnetic field is applied on the magnetic texture. We show a simulation for $N_{tot}$ = 60, $N_1$ = 40, $N_2$ = 20, t = 100, $\Delta = 1, t_s = 1, \Gamma_N = 0, \gamma_n = 0.1, \mu = -0.85 * 2t, B_{osc} = 1$ .

We now investigate the effect of disorder in the oscillating magnetic field, using the tight-binding model introduced above. We take two models for the disorder. Extended data Fig. 13 illustrates the result of this study.

First, we consider a field that evolves in space in the same fashion as the MFM cut of Fig. 1. This signal contains various frequencies, as shown by the discrete Fourier transform given in Extended data Fig. 7b. We construct a cycloidal magnetic field from the Fourier coefficients $a_f$ associated with the spatial frequency f of the MFM signal in the following way:

$B_{osc,z}(i) = B_{osc} \sum_f a_f \sin(2\pi f i)\,, B_{osc,x}(i) = \sum_f \mathrm{a_f} \cos(2\pi f i))$ .

We then model its evolution under an external magnetic field by a simple shift of the frequency of each coefficient of the Fourier transform. The effect of $B_{ext}$ is:

$$a_f(B_{ext}) = a_{f+\delta f(B_{ext})}(3)$$

With $\delta f(B_{ext}) \propto B_{ext}$.

We use as the parameters for the discrete Hamiltonian: $N_{tot}$ = 60, $N_1$ = 40, $N_2$ = 20, t=100, $\Delta = 1, t_s = 1, \Gamma_N = 0.2, \gamma_n = 0.15, \mu = -0.82 * 2t$. The amplitude of the oscillating field $B_{osc}$ is normalized at each $B_{ext}$ such that the maximal amplitude is 1 (in units of the superconducting gap Δ). We then plot the density of states at the first site of the chain, as a function of energy and $B_{ext}$. If we consider a situation where there is only one oscillation frequency, we obtain a very similar result, namely oscillations of a pair of ALSs. We conclude that disorder in the magnetic field does not strongly affect the oscillations of the ALSs predicted for a periodic magnetic field, and observed in the experiment.

As an alternative approach, we use the magnetic simulations of section 5 from which we can directly extract the magnetic field in all three directions above the Co/Pt structure, and its evolution as a function of the external magnetic field. This is shown in Extended data Fig. 7f. We use as the parameters for the discrete Hamiltonian: $N_{tot}$ = 60, $N_1$ = 40, $N_2 = 20$, t=100, $\Delta = 1, t_s = 1, \Gamma_N = 0, \gamma_n = 0.15, \mu = -0.85 * 2t$. The amplitude of the oscillating field is taken as:

$B_{osc}$ (units of $\Delta$) $= 2\ B_{magnetic\ simulations}$(T)

in order to have a qualitative agreement with the measured oscillations, for this set of parameters.

The micro-magnetic simulations give the spatial evolution of the field vector $B_{osc}$ for seven values of $B_{ext}$ : 0, 0.2, 0.4, 0.6, 0.8, 1 and 1.2T. We interpolate linearly the evolution of each component of the field in between these seven values. We plot the density of states at the first site as a function of energy and $B_{ext}$, and observe oscillations of the ALSs energies, under this more realistic evolution of the magnetic texture. The simulations reproduce qualitatively the evolution of $B_{osc}$ in the three direction of space, when a magnetic field is applied to the magnetic texture, and at different cut positions we can obtain different realizations for $B_{osc}$. The simulated fields used for this study may not perfectly fit our sample's stray field; notably it seems to contain more domains than what the MFM signal indicates.

To conclude, this show that the analysis of the oscillations of the ALSs is robust considering more realistic models of our system, build either from a realistic stray field profile extracted from the MFM data, or from micro-magnetic simulations.

We now turn to the situation where Majorana zero modes can potentially emerge in our setup. Our setup is different from the ones which are a priori used in semiconducting systems. We show in Extended data Fig. 14 the relevant figures. We study the variations of the density of states (DOS), the singlet pairing and the triplet pairing as a function of the number of magnetic domains (which are tuned by the external magnetic field in the experiment). The parameters of the model are given below

$N_{tot}$ = 20, $N_1$ = 16, $N_2$ = 15, t=12, $\Delta = 1, t_s = 12, \Gamma_N = 0.1, \gamma_n = 0.05, \mu = -0.33 * 2t, B_{osc} = 8$.

The density of states shown in Extended data Fig. 14a displays oscillations of Andreev Like states at non zero energy but more importantly a zero bias peak emerging when the number of domains increases. In order to characterize the Andreev Like states, it is instructive to plot the singlet and triplet pairing amplitudes (in panel b and c) for the same parameters. Interestingly, before the emergence of the zero bias peak, one sees both singlet and triplet pairing amplitudes as expected for superconductivity in the presence of homogeneous spin polarization. However, the emerging zero bias peak is solely made out of triplet correlations, as required for a Majorana zero mode. The singlet pairing amplitude is defined as $\frac{\left|G^R\left(d_\uparrow^\dagger, d_\downarrow^\dagger\right) - G^R\left(d_\downarrow^\dagger, d_\uparrow^\dagger\right)\right|}{\sqrt{2}}$, where G is the retarded Green's function, and e (resp. h) refers to electrons (resp. holes). The triplet pairing amplitude is defined as

$$\sqrt{\frac{1}{2}\left|G^R\left(d_\uparrow^\dagger, d_\downarrow^\dagger\right) + G^R\left(d_\downarrow^\dagger, d_\uparrow^\dagger\right)\right|^2 + \left|G^R\left(d_\uparrow^\dagger, d_\uparrow^\dagger\right)\right|^2 + \left|G^R\left(d_\downarrow^\dagger, d_\downarrow^\dagger\right)\right|^2}.$$

We now show another important feature of this zero mode: its spatial localization and sensitivity to the local configuration of the magnetic field, as illustrated in panels e, f and g. These colormaps show the density of state as a function of the position in the chain and energy, for a system with the same parameters as panel a and with 7 magnetic field oscillations (7 domains). As expected, we see that the zero bias peak corresponds to two localized states located at the interface between the magnetic texture and the superconductor and the magnetic texture and the left hard wall. On the contrary, the non-zero energy Andreev Like States are fully delocalized on the entire wire length.

Importantly a change in the magnetic texture has barely any effect on the later irrespectively of the position of this change (Extended data Fig. 14e and f). This is completely different from the case of the Majorana zero mode which is insensitive to a local reconfiguration of the magnetic texture if this reconfiguration occurs in the middle of the wire (Extended data Fig. 14g) whereas it splits if the reconfiguration occurs close to the superconducting/wire interface where its wave function is non-zero (Extended data Fig. 14f).

All these features reproduce qualitatively our experimental findings and show that localized Majorana zero modes can emerge when superconductivity is induced from the side of the wire, in a different manner than the initial propositions for engineering Majorana bound states in one dimensional conductor.

In summary, our numerical study confirm the robustness to disorder of the oscillations of the ALSs at non-zero energy and substantiate the Majorana zero mode interpretation of the observed zero bias peak.

**Data availability.** The authors declare that the main data supporting the findings of this study are available within the article (main text, methods and extended data). Extra data are available from the corresponding author upon request.

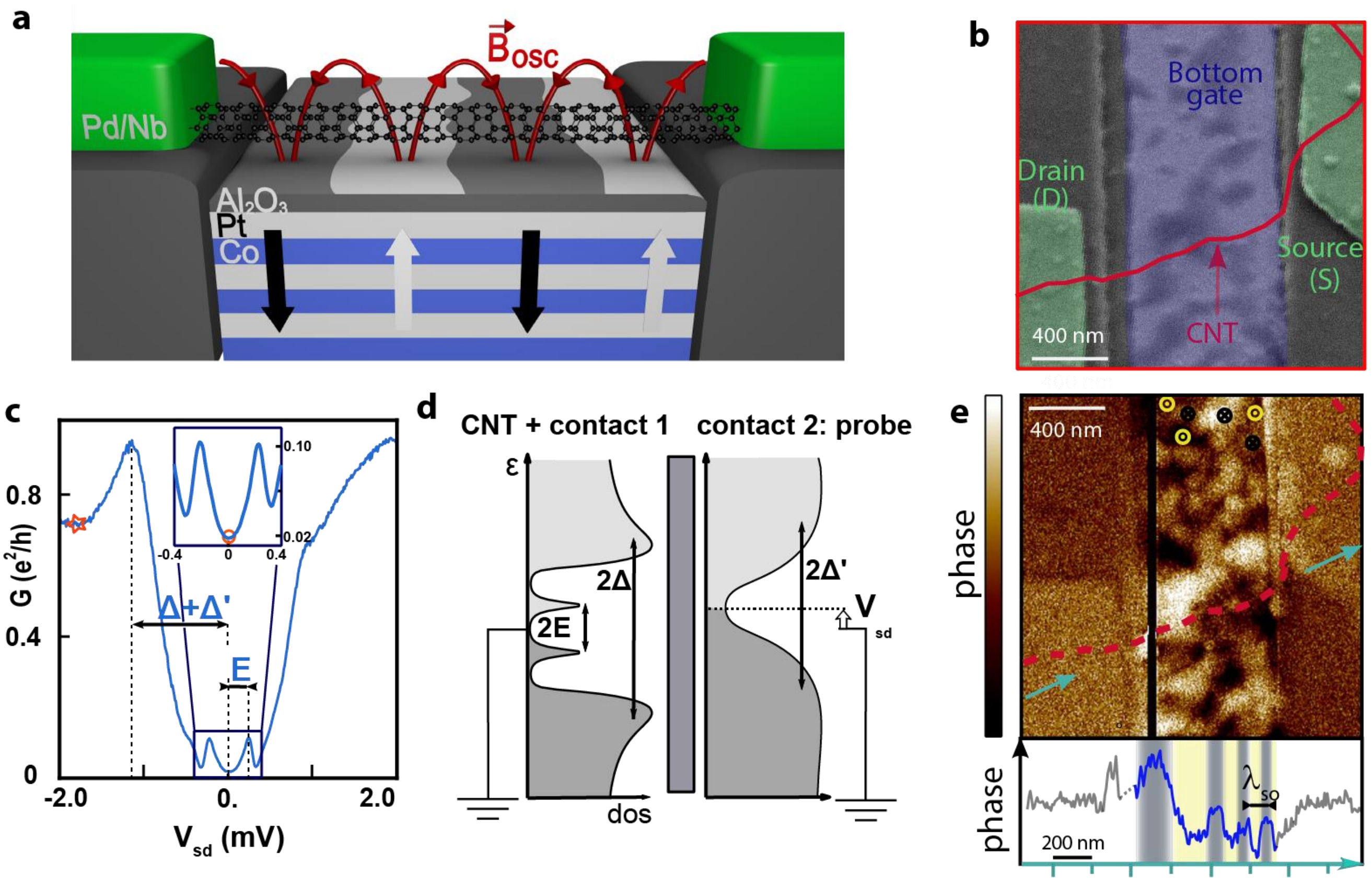

Fig. 1 Desjardins et al.

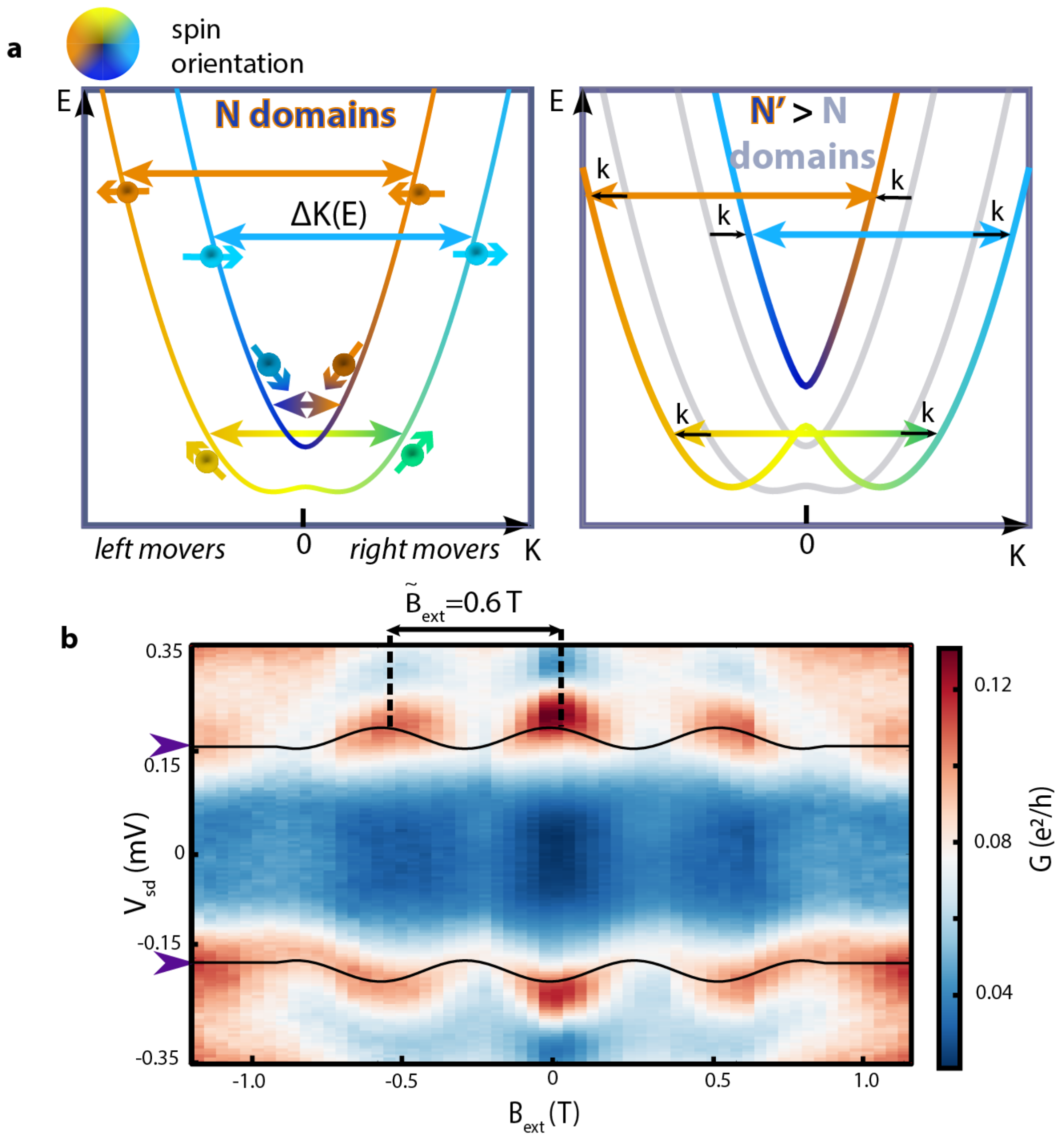

Fig. 2 Desjardins et al.

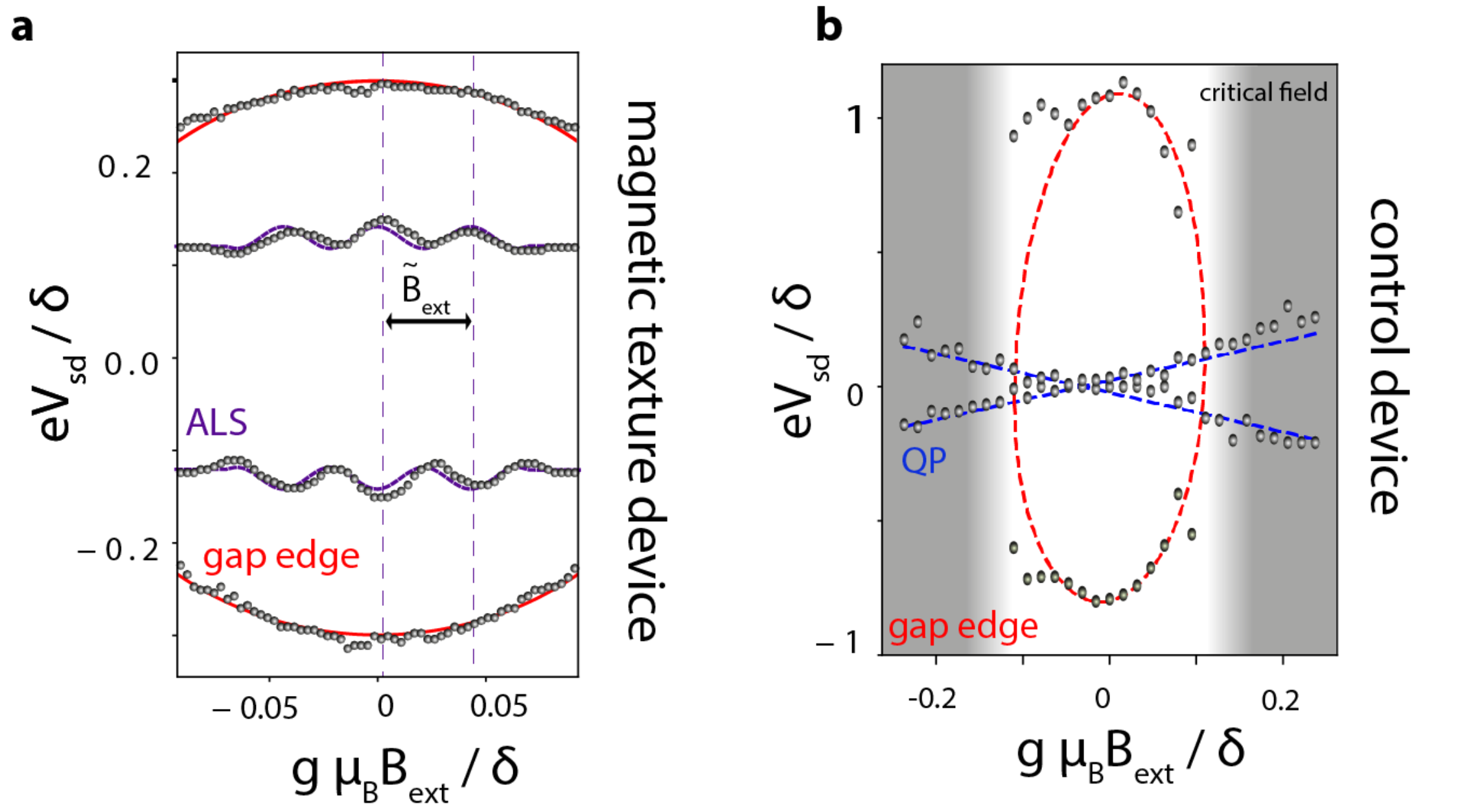

c

| scenario | ALS | | | QP | gap |
|---|---|---|---|---|---|
| | $E_{so} = 0$ | $E_{so}$ constant | $E_{so}(B)$ | | |
| evolution in $B_{ext}$ | oscillations crossing at 0 | oscillations | oscillations | splitting | closing |
| characteristic field | $\tilde{B}_{ext} = \delta / g\mu_B$ | $\tilde{B}_{ext} > \delta / g\mu_B$ | $E_{so}(\tilde{B}_{ext}) = E_{so}(0) - \delta$ | $X$ | $B_c$ |
| order of magnitude | $\tilde{B}_{ext} \sim 5T$ | $\tilde{B}_{ext} > 5T$ | depends on $E_{so}(B_{ext})$ | $X$ | $1 - 2T$ |

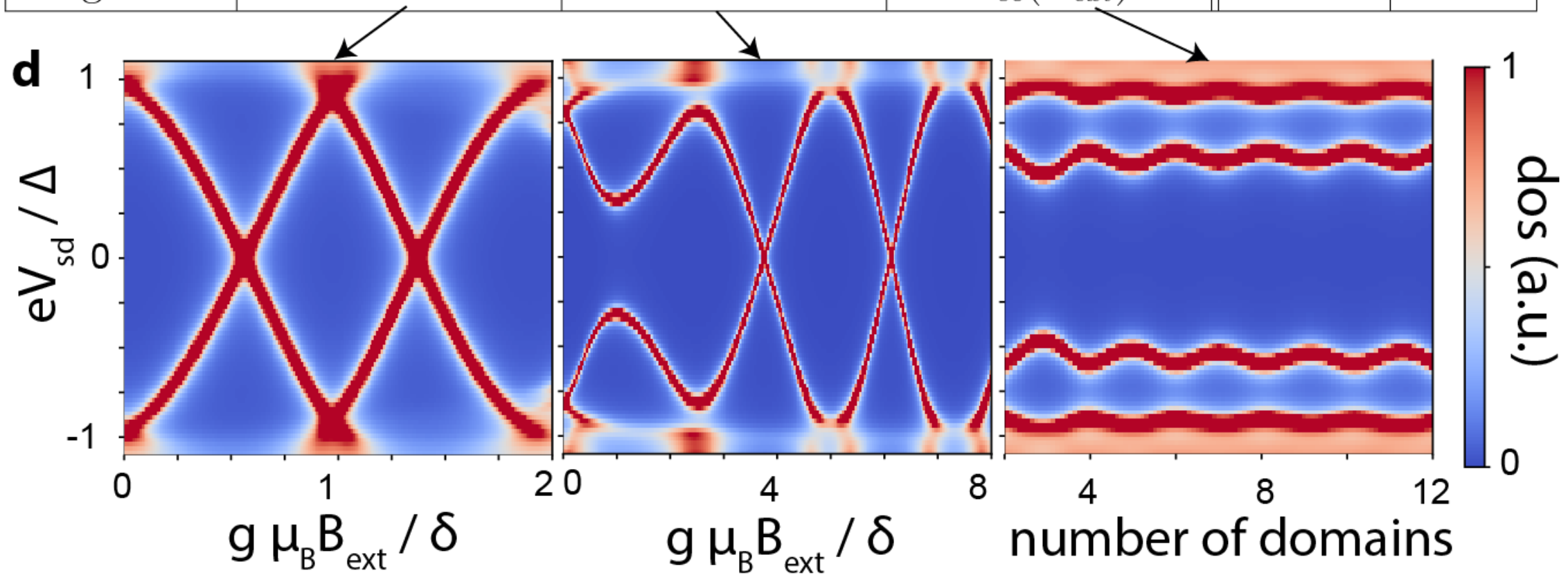

Fig. 3 Desjardins et al.

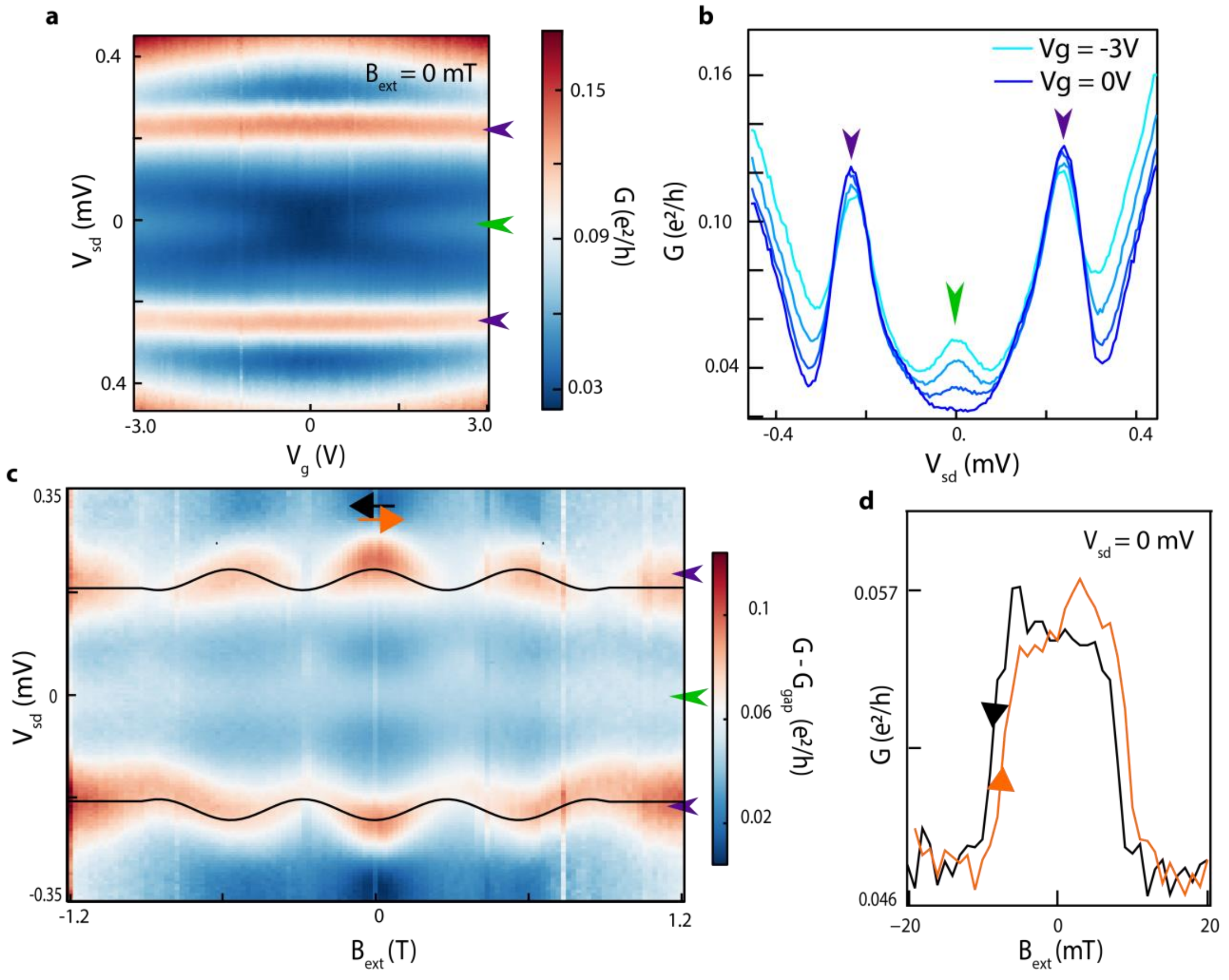

Fig. 4 Desjardins et al.

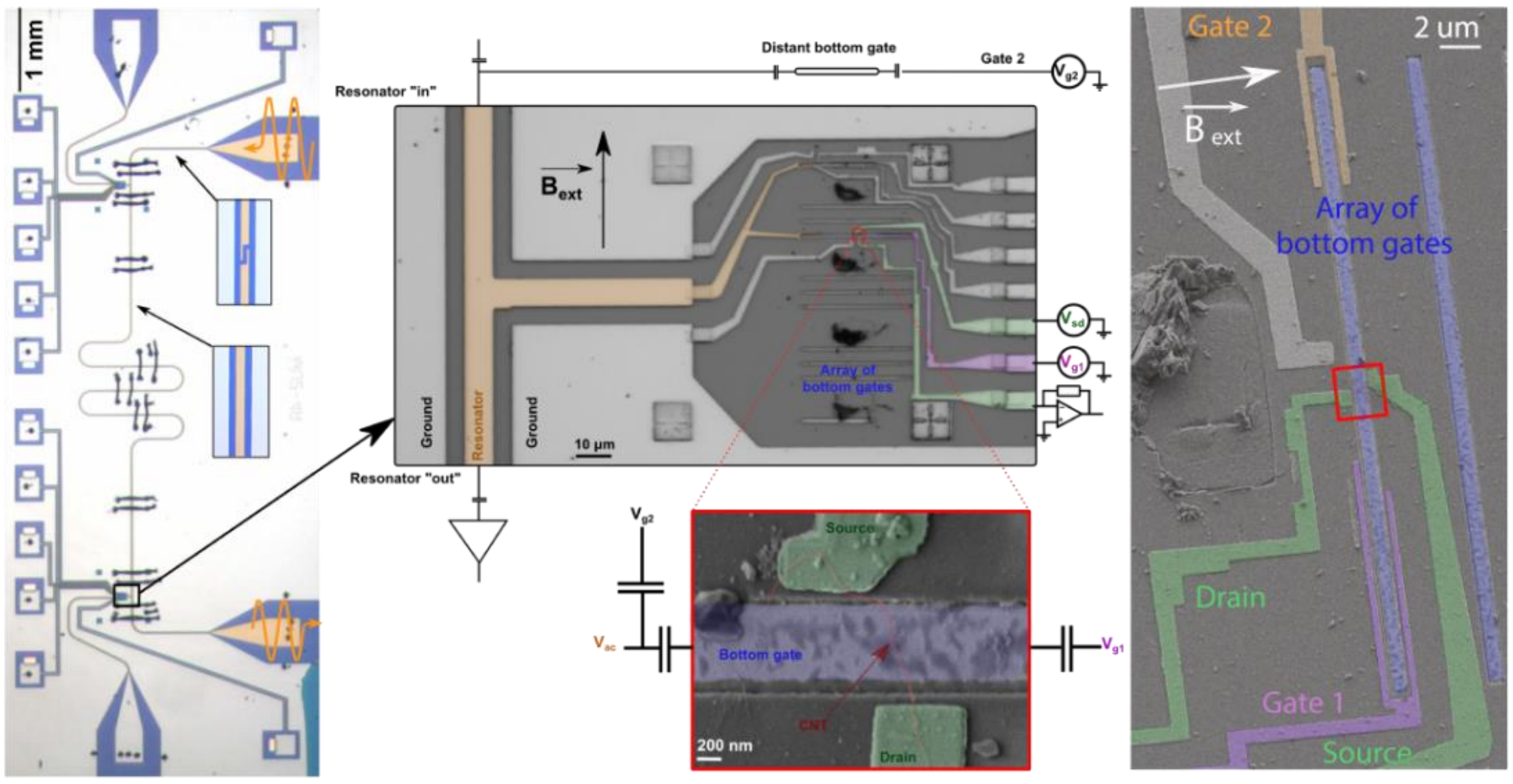

Extended Data Fig. 1
Desjardins et al.

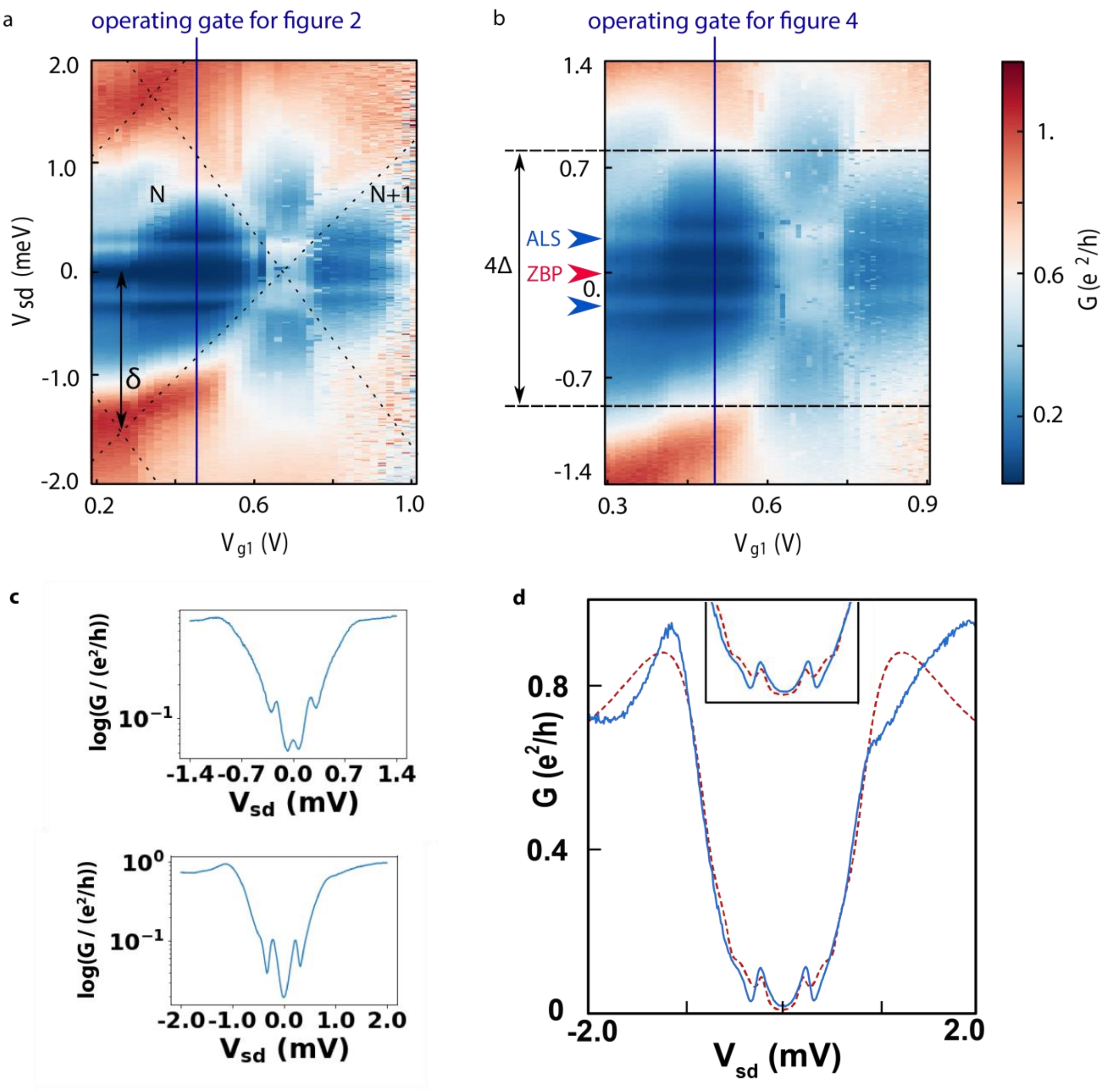

Extended Data Fig. 2
Desjardins et al.

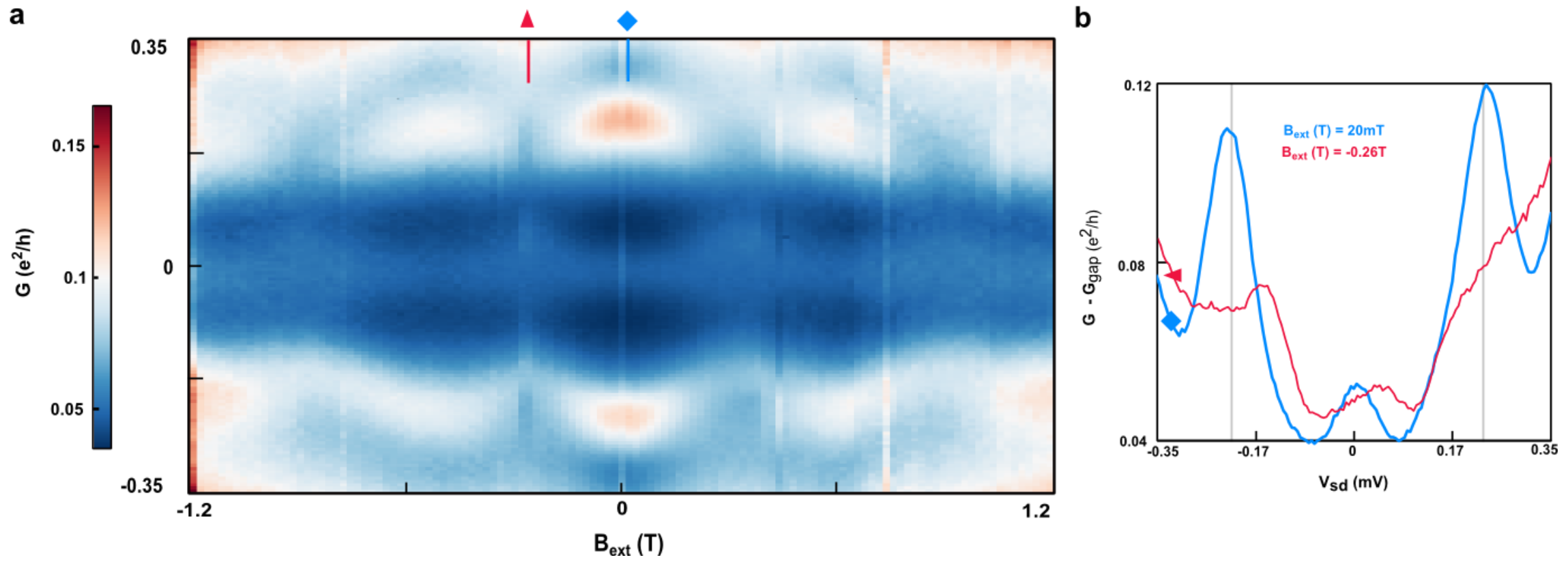

Extended Data Fig. 3
Desjardins et al.

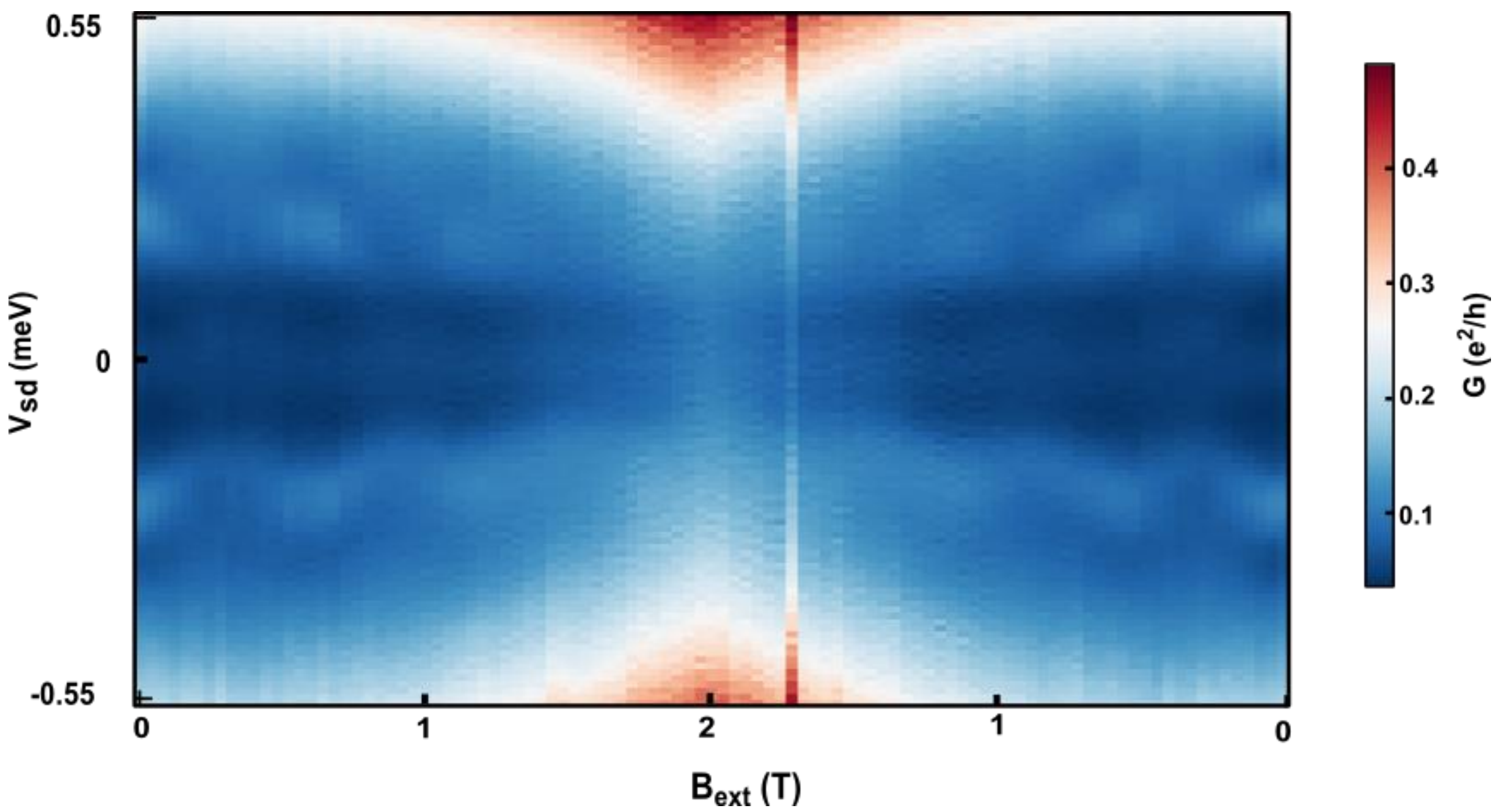

Extended Data Fig. 4
Desjardins et al.

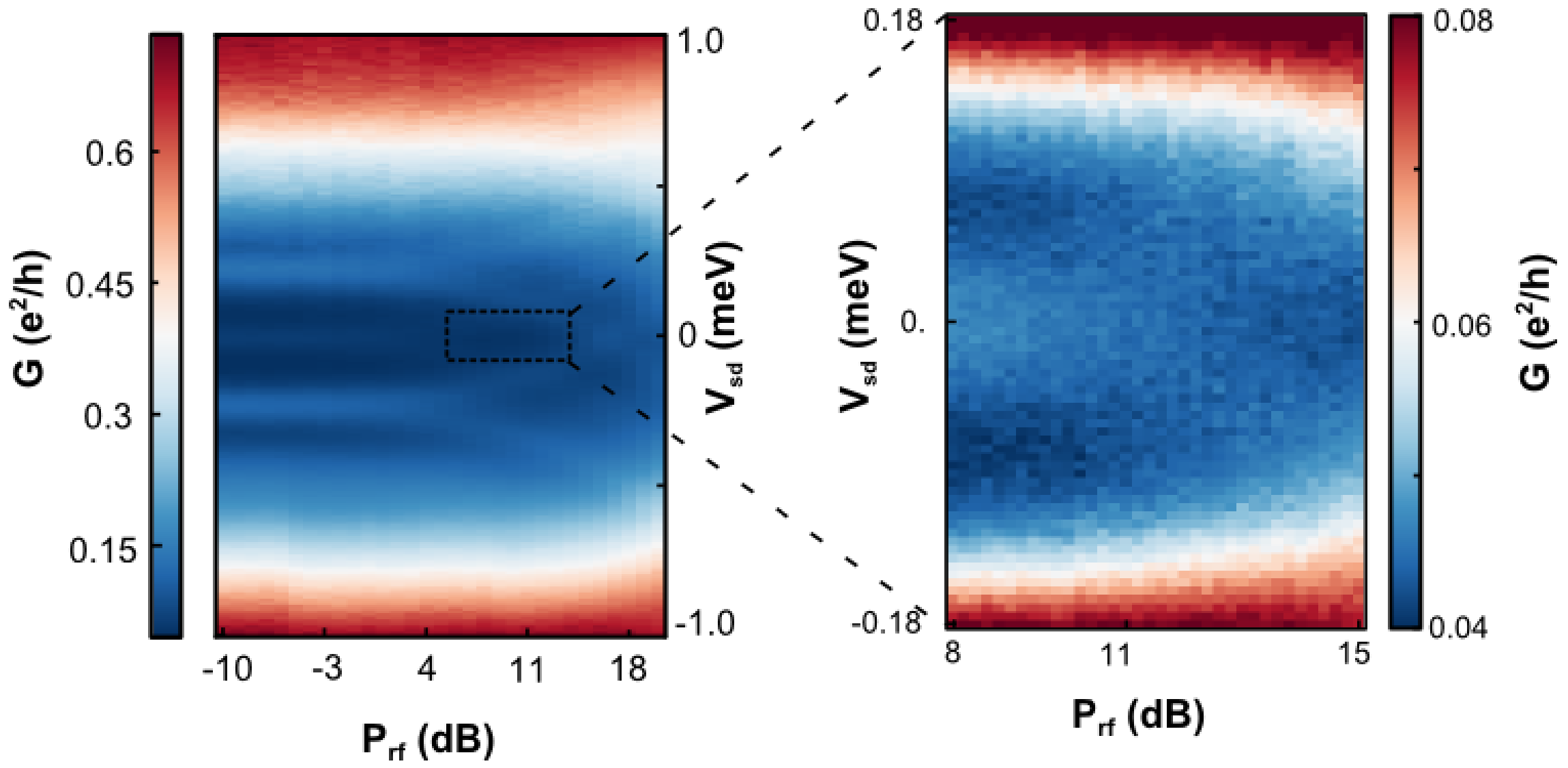

Extended Data Fig. 5
Desjardins et al.

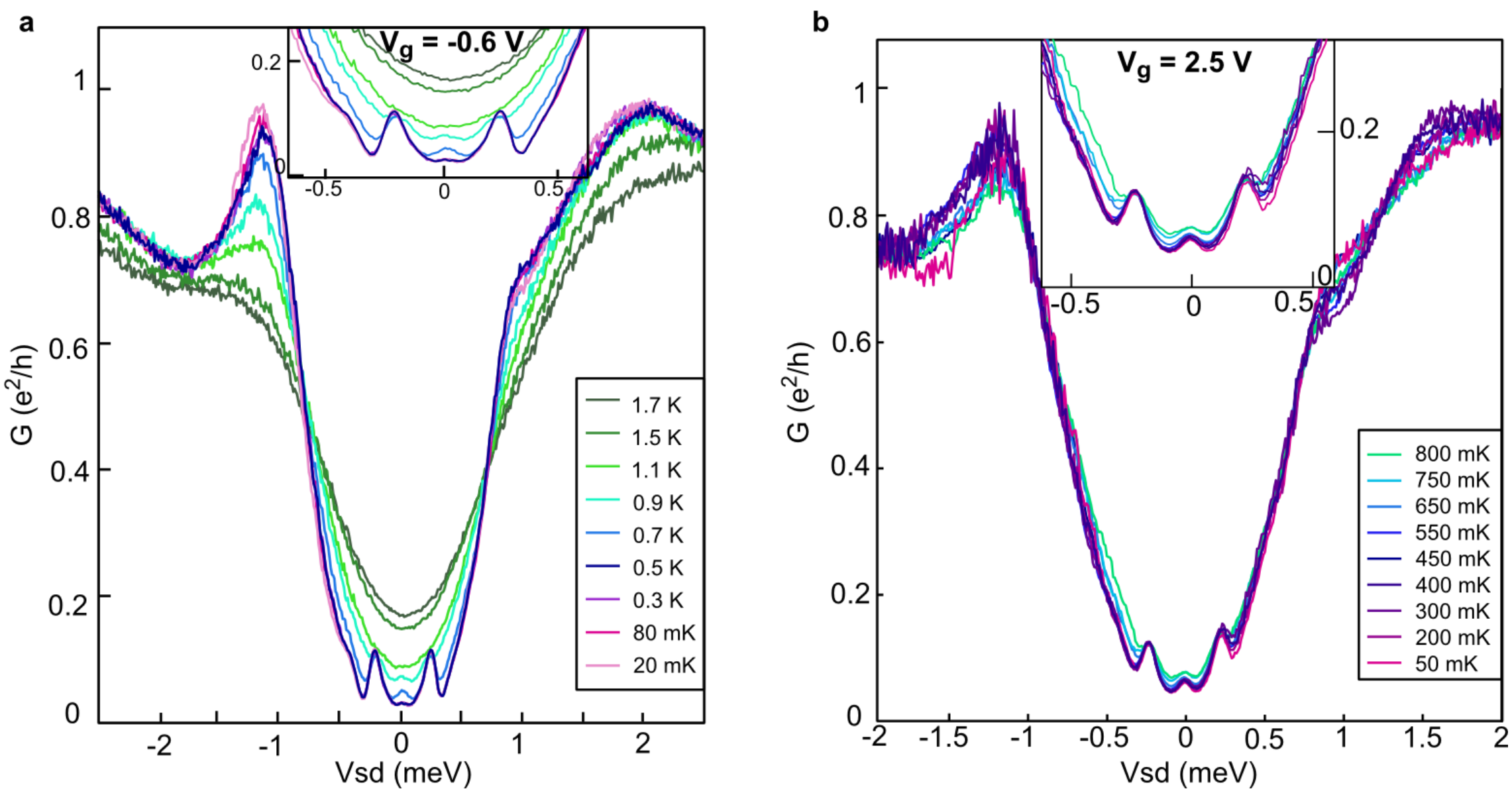

Extended Data Fig. 6
Desjardins et al.

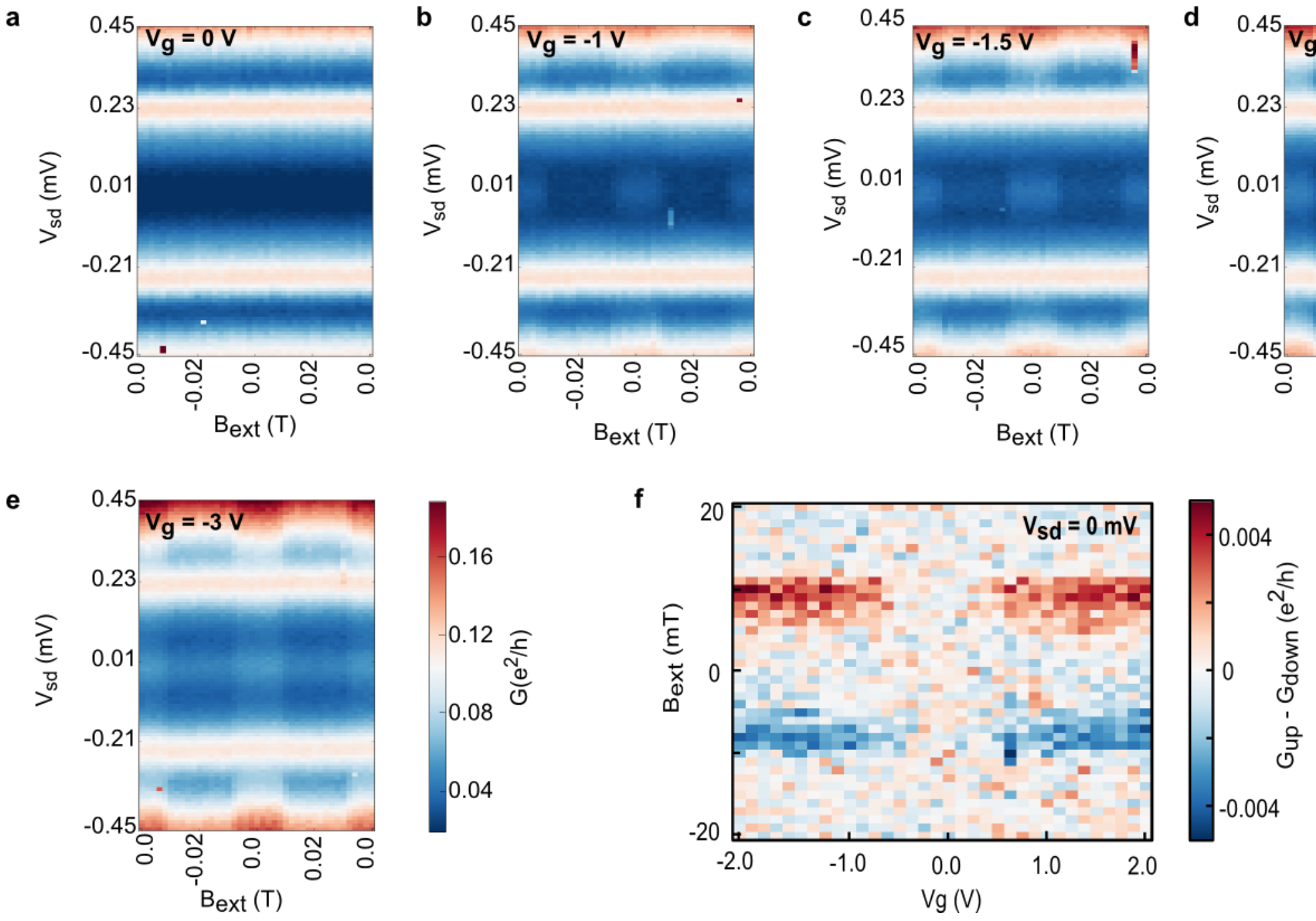

Extended Data Fig. 7
Desjardins et al.

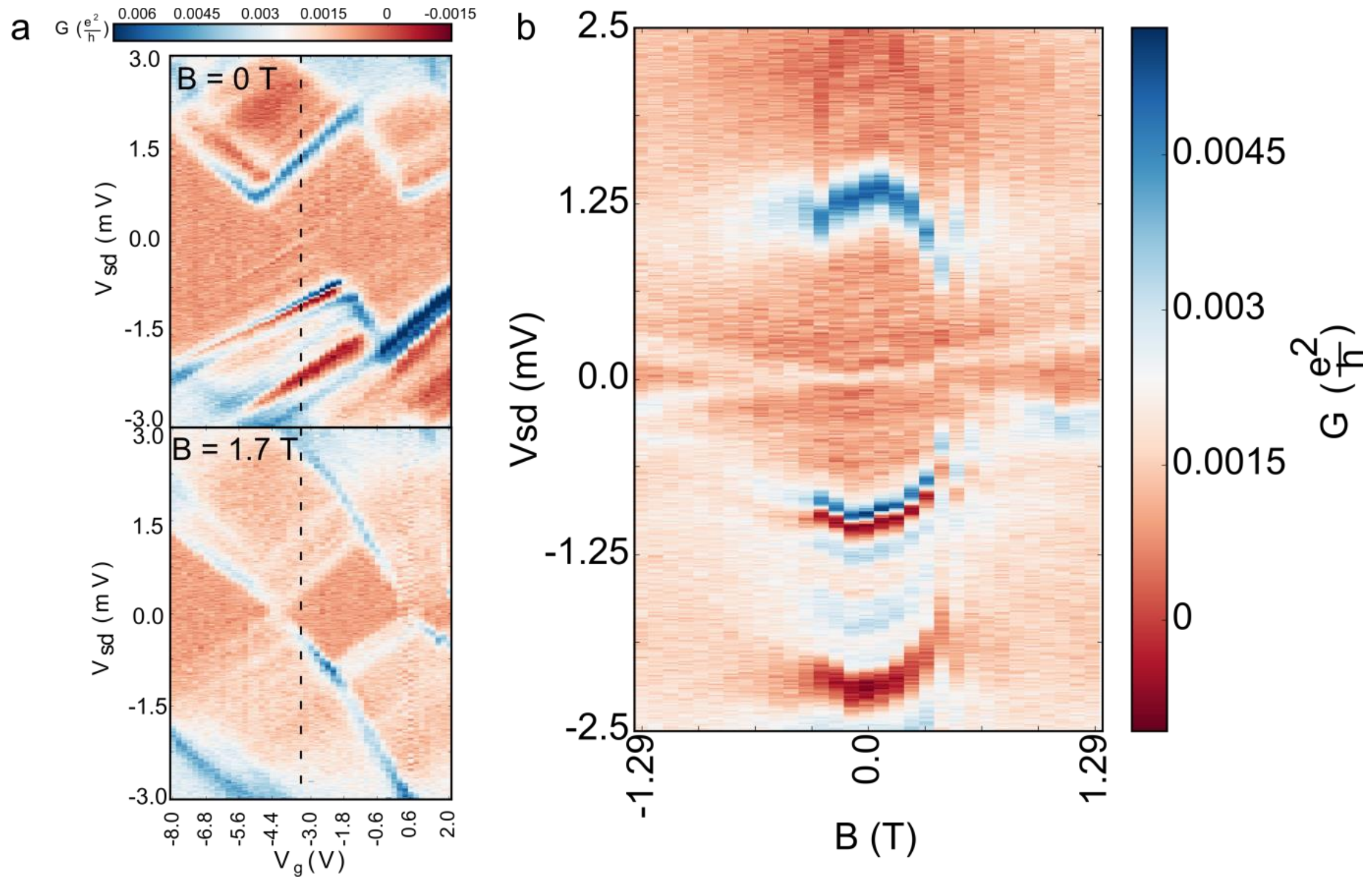

Extended Data Fig. 8
Desjardins et al.

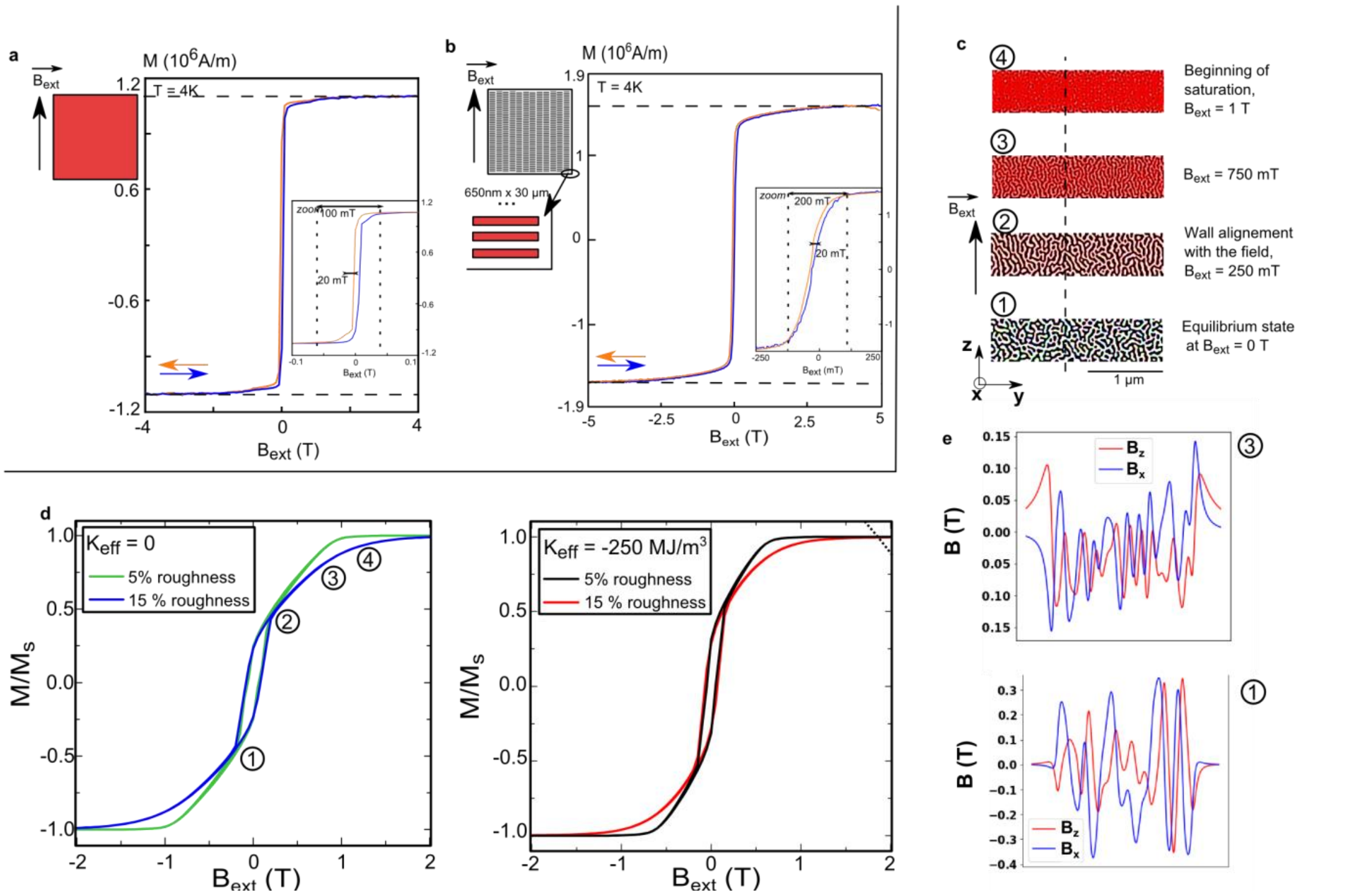

Extended Data Fig. 9
Desjardins et al.

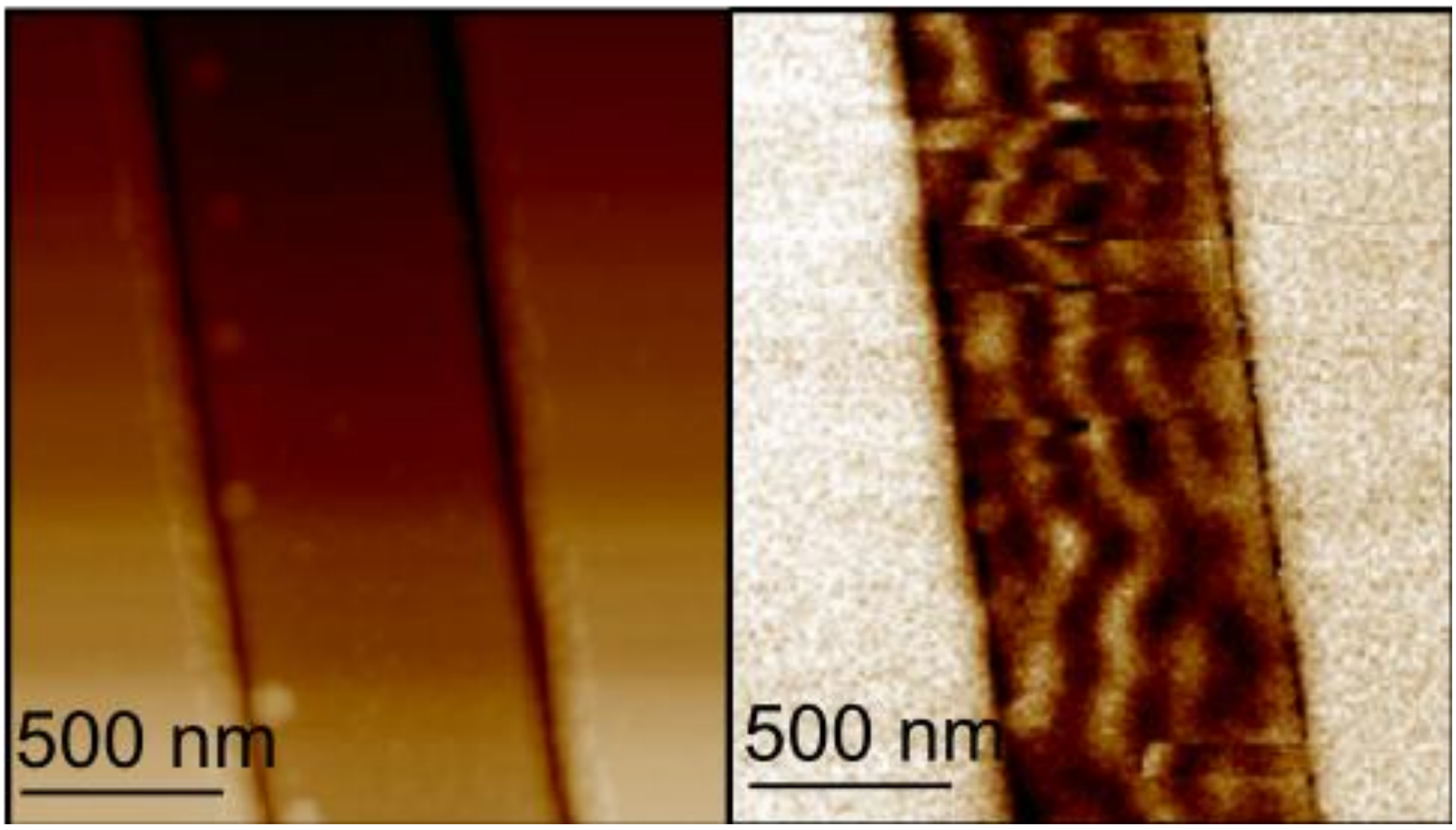

Extended Data Fig. 10
Desjardins et al.

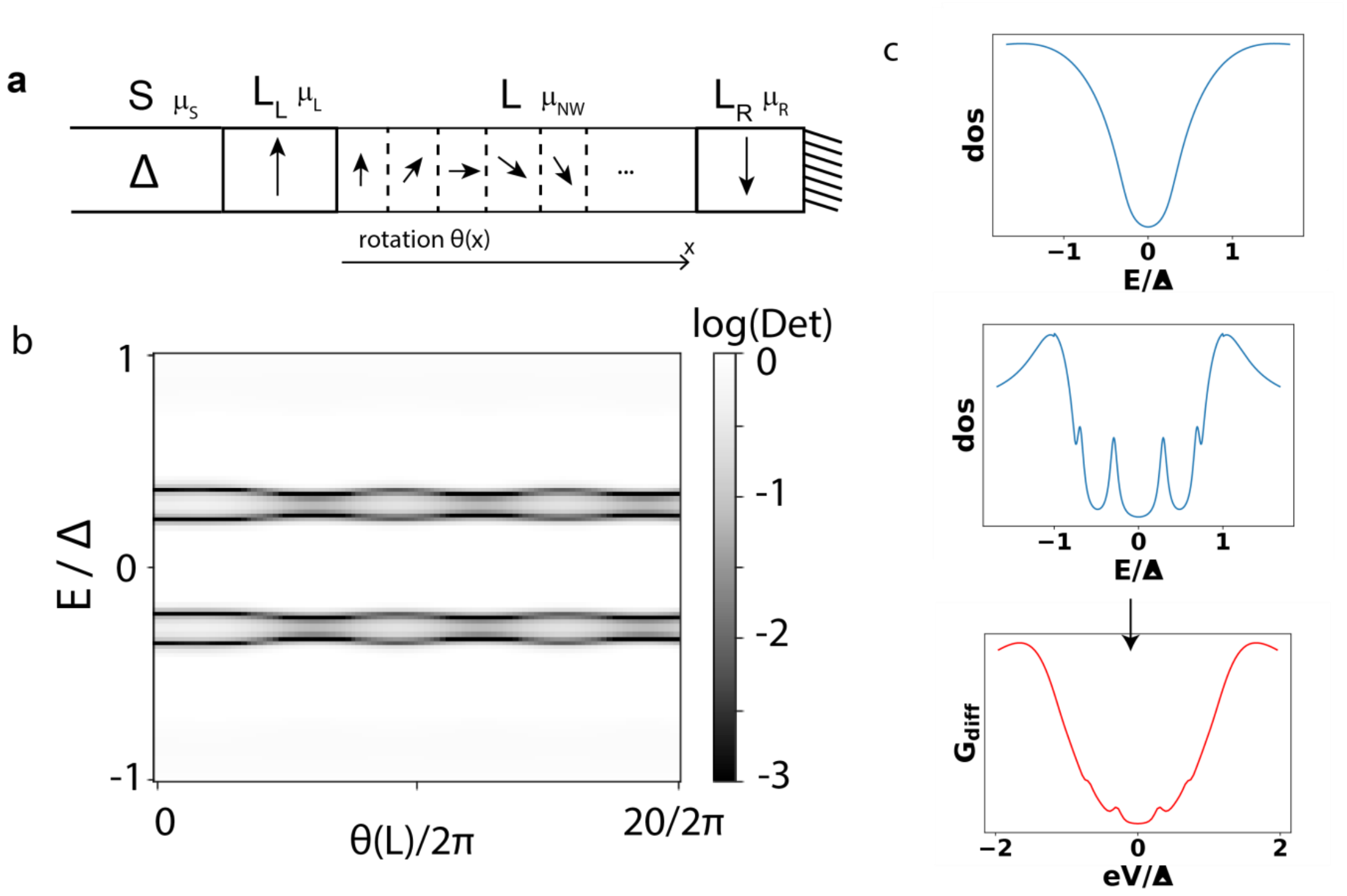

Extended Data Fig. 11
Desjardins et al.

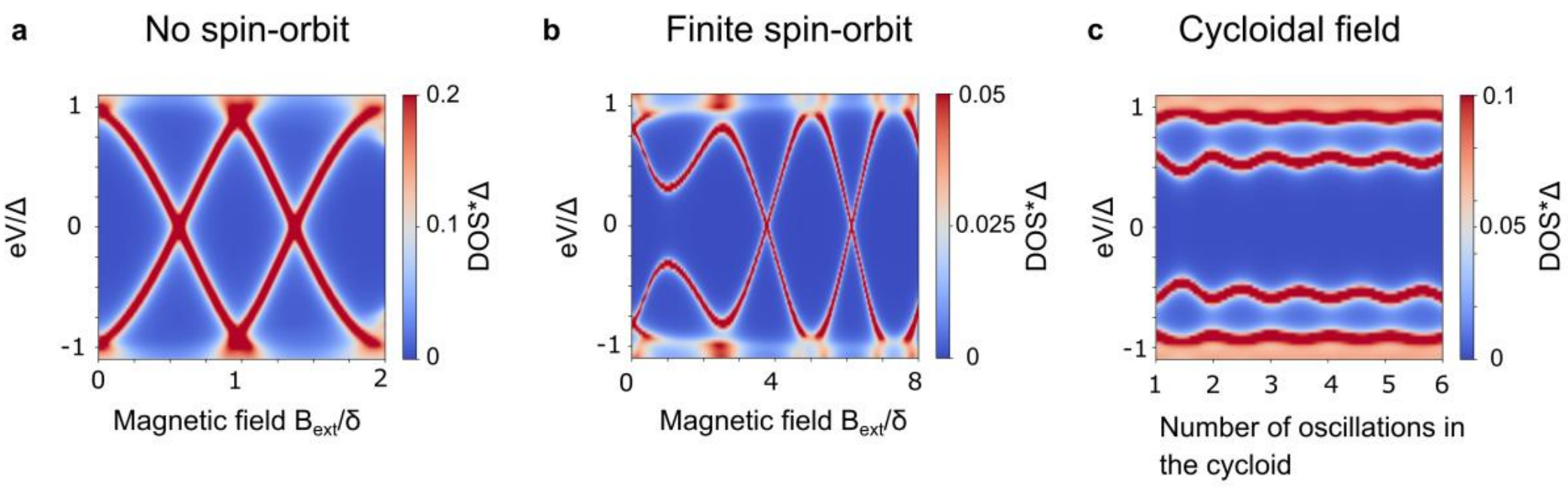

Extended Data Fig. 12
Desjardins et al.

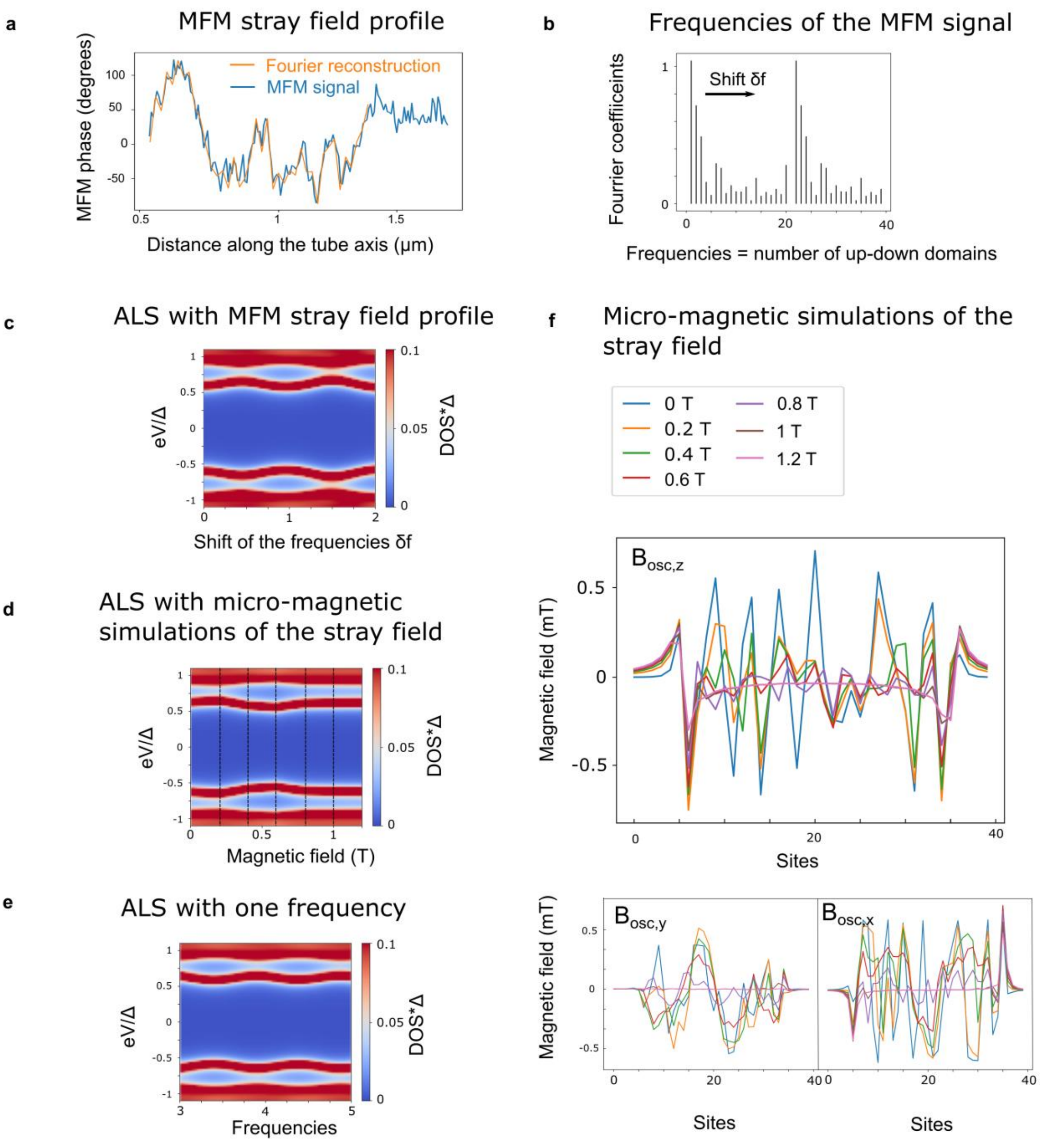

Extended Data Fig. 13
Desjardins et al.

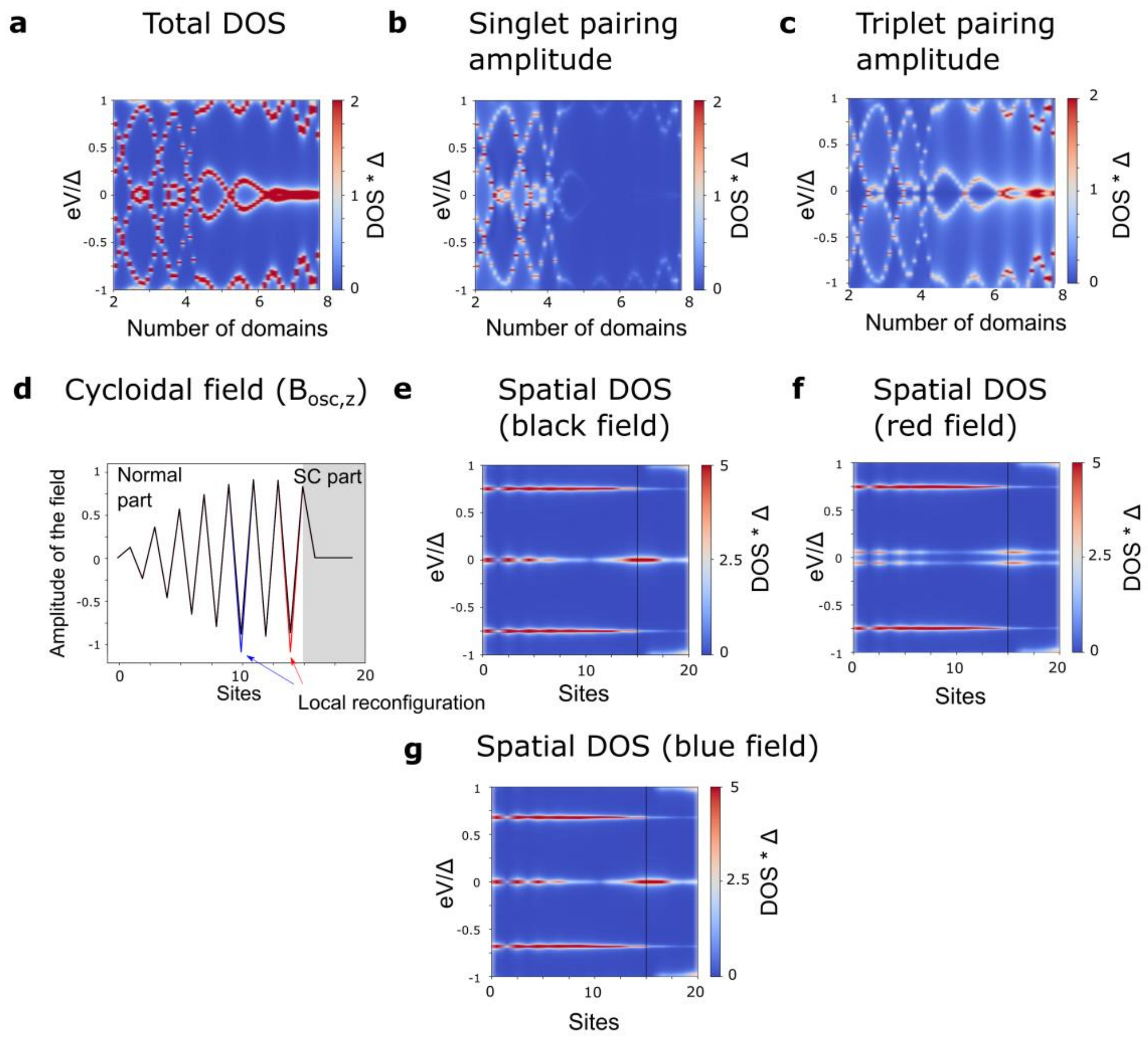

Extended Data Fig. 14
Desjardins et al.